\documentclass[aps,prx,reprint,superscriptaddress,longbibliography,twocolumn]{revtex4-2}
\usepackage{graphicx} 
\usepackage{amsmath}
\usepackage{amssymb}
\usepackage{color}
\usepackage[colorlinks=true,linkcolor=blue, citecolor=green, urlcolor=blue]{hyperref}
\usepackage{multirow,booktabs}

\definecolor{eggplant}{RGB}{126,93,181}
\definecolor{cayenne}{RGB}{148,17,0}
\definecolor{teal}{RGB}{0,145,147}
\definecolor{blueberry}{RGB}{4,51,255}

\newcommand{\ket}[1]{| #1 \rangle}
\newcommand{\bra}[1]{\langle #1 |}

\begin{document}

\title{Process Tensor Approaches to Non-Markovian Quantum Dynamics}
\date{December 2023}
\author{Jonathan Keeling}
\affiliation{SUPA, School of Physics and Astronomy, University of St Andrews, St Andrews, KY16 9SS, United Kingdom}
\author{E. Miles Stoudenmire}
\affiliation{Center for Computational Quantum Physics, Flatiron Institute, 162 Fifth Avenue, New York, NY 10010, USA}
\author{Mari-Carmen Ba\~nuls}
\affiliation{Max-Planck-Institut für Quantenoptik, Hans-Kopfermann-Straße 1, D-85748 Garching, Germany}
\affiliation{Munich Center for Quantum Science and Technology, Schellingstraße 4, 80799 München, Germany}
\author{David R. Reichman}
\affiliation{Department of Chemistry, Columbia University, New York, NY 10027, USA}

\begin{abstract}
    The paradigm of considering open quantum systems---i.e.\ focusing only on the system of interest, and treating the rest of the world as an effective environment---has proven to be a highly effective way to understand a range of quantum systems, across areas of study such as quantum optics, cold atoms, superconducting qubits, and impurities in solid-state systems.  
    A common approach in many of these contexts has been to consider simplified approaches based on the Born and Markov approximations. 
    While these approximations are indeed often appropriate in contexts such as quantum optics, the widespread application of these approximations has been driven more by simplicity than by accuracy.  
    In particular, these Markovian treatments will fail in many cases, such as when coupling to the environment is not weak, when the environment is structured and has resonances, when the system couples to low-frequency modes of the environment, or when the questions of interest involve the propagation of information through the environment.
    Despite the fact that many real problems are non-Markovian, the Markov approximation is still widely used, as it is often assumed that a fully non-Markovian treatment is too complex to be practical.
    In this perspective we discuss a recently developed set of techniques that address this challenge. Centering our discussion around the notion of the process tensor, we demonstrate that the generality of the process tensor concept, coupled with efficient tensor-network methods, opens the door to the description of a wide range of observable non-Markovian processes in a wide range of open quantum systems.
\end{abstract}

\maketitle


\section{Introduction}

Although quantum mechanics is often first introduced to students by focusing on idealized isolated quantum systems, no real-world quantum system is truly isolated, particularly not those that can be manipulated or measured. 
As such, understanding how quantum systems interact with their environment is an important task and one which has been extensively studied~\cite{breuer2002:theory}.
As discussed below, much work on open quantum systems has focused on particular cases, where a simple time-local---i.e. Markovian---description can apply \footnote{Note here by time-local we exclude situations where memory effects may be hidden by the formalism.  For example, the time-convolutionless Master equation framework~\cite{breuer2002:theory} appears to be local in time, but the time-dependent relaxation term depends in principle on the entire bath history starting from the initial time}.
However, as also discussed below, many open quantum systems do not warrant this Markovian approximation~\cite{DeVega2017Dynamics}.
The process tensor, which we will introduce in the following, is an object that allows one to predict the outcome of a sequence of operations in a general quantum system coupled to an environment~\cite{Chiribella2008,Chiribella2009,Pollock2018NonMarkovian,Milz2021Quantum,Taranto2025HigherOrder}.
As such, if known, the process tensor can be used to model general non-Markovian open quantum systems.

This perspective discusses recent developments in using the process tensor framework, and particularly tensor network representations~\cite{Verstraete2008matrix, Schollwoeck2011DensityMatrix, Orus2014Practical,Bridgeman_2017, Silvi2019tensor,Ran2020tncontr,Cirac2021rmp,Okunishi2022,Banuls2023Tensor} of the process tensor, to enable practical calculations of non-Markovian open quantum systems.
We will discuss how the process tensor framework provides both a numerical tool that has expanded the range of such problems which can be efficiently modeled, as well as  a way to  unify many previous algorithms that have been used for non-Markovian open quantum systems.
Moreover, we will discuss ways in which this connects to previously distinct problems in many-body physics, such as methods for modeling impurities, and highlight future research directions in this field.  
The ideas we discuss here historically developed from previous approaches based on tensor-network representations of the augmented density tensor from the quasi-adiabatic propagator path integral (QuAPI) approach~\cite{Makri1995Tensor1,Makri1995Tensor2}. 
We will introduce the process tensor and its tensor network representation from a different perspective, with the aim of distinguishing the key ideas of this approach from their historical development. Specifically, as we discuss below, while process tensor approaches can be derived from the QuAPI approach, they do not have to be derived this way.

In order to set the scene for such a discussion, the rest of this introduction will discuss two connected themes:  non-Markovian open quantum systems---and the wide range of contexts in which the open quantum system paradigm applies---and tensor network methods as a route to dimensional reduction.

\subsection{The importance of non-Markovian open quantum systems}

The open quantum system (OQS) approach is designed to simplify the modeling of a complex quantum system by dividing it into the ``system'' and the ``environment.''
The goal is to be able to write a description of the system, accounting for effects of the environment, without explicitly modeling its many degrees of freedom.
Historically, much of the theory of open quantum systems  was first developed in the context of quantum optics, to describe effects such as dissipation due to radiation.
In this context, where linewidths are much smaller than frequencies, the Markovian approximation holds very well.
As such, extensive work has been done to develop the theory of Markovian open quantum systems, so this subject is well described in textbooks~\cite{Scully1997:qo,breuer2002:theory} and calculations within it are routine through standard libraries such as QuTip~\cite{Johansson2012Qutip,Johansson2013Qutip2} and QuantumOptics.jl~\cite{kramer2018quantumoptics}.
In contrast, the simulation of non-Markovian systems has relied on more specialized methods until recently, making them less accessible to a wide user base.

This history can give the impression that the Markovian approach is ``default'' and that treatment of non-Markovian dynamics is a ``niche'' topic.
This impression is reinforced by the view that, in cases where the approximations and assumptions required for the standard textbook approach fail, this indicates a failing of the OQS paradigm itself.
That is, one may have the impression that in cases where weak coupling or Markovianity is not applicable,  one cannot eliminate the environment, and one must resort to methods which attempt to explicitly include environmental degrees of freedom within the system.
As we discuss below, this need not be the case---particularly with numerical methods that broaden the range of environments that can be accurately described.

Much has been written about the different ways in which Markovianity, and thus non-Markovianity, can be formally defined~\cite{Rivas2014Quantum}.
For the questions we discuss in this perspective, such distinctions are not generally significant in that the motivation here is to model systems in which standard Markovian approaches such as Lindblad master equations fail, ultimately due to memory effects in the environment or correlations and entanglement between the system and environment.
In fact, as discussed elsewhere~\cite{Pollock2018Operational}, the tensor network representation of the process tensor can be used as a definition of (non-)Markovianity.
As such, the tensor network process tensor paradigm naturally becomes increasingly simple to calculate in those cases where a system is Markovian, making it unnecessary to determine {\em a priori} whether the system is or is not Markovian.   In that sense, while the title of this perspective refers to ``non-Markovian'' dynamics, this should be read as meaning ``not necessarily Markovian'' throughout.

\subsubsection{The ubiquity of non-Markovian open quantum systems}

The open quantum system paradigm that considers the system as separate from its environment does not rely on the ability to ``adiabatically eliminate'' the environment, or any assumption of weak coupling between system and environment.   
The only crucial feature required for the OQS paradigm is that one is interested \emph{only} in \emph{measurements on the system}.  
If that is true, the paradigm holds, and there will be some (generically non-Markovian) treatment that can accurately capture the physical behavior.   

This idea is widely appreciated in the context of dynamical mean field theory (DMFT), in which a key ingredient is the need to solve the dynamics of a quantum ``impurity'' in a general environment, with no expectation that this problem should be Markovian~\cite{Georges1996Dynamical} or weakly coupled.
The same methods that can be used to tackle non-Markovian open quantum systems can be considered as impurity solvers for DMFT~\cite{Thoeniss2023Efficient,Ng2023Realtime,Sonner2025}---indeed, DMFT can be seen as an application of the OQS paradigm to many-body physics.

With the perspective outlined above, non-Markovian open quantum system methods have a far wider applicability than is typically considered in discussions of OQS.  They are in fact relevant in any problem where (a) one is interested only in measurements on some subset of degrees of freedom, and (b) one cannot use weak-coupling and Markovian approximations to separate out those degrees of freedom as a system weakly coupled to an environment.  
This can include many-body physics in contexts such as impurity and transport problems, or contexts where one studies local probes of correlated electronic materials, such as spectroscopic scanning tunnelling microscopy.
The reason these are not commonly thought of as examples of non-Markovian open quantum systems is that finding methods to efficiently model them as such can be challenging.

\subsubsection{Limitations of Markovian methods}

While the discussion above aims to broaden the context of where non-Markovian OQS problems may arise, we next discuss cases which are perhaps more familiar to consider as open quantum systems.  Within this context,  there are broadly two categories of reasons why a Markovian approach may be inadequate, as discussed next.

\paragraph{The system does not permit it.}
This category includes any problem where the standard approximations required to derive a Markovian master equation break down---specifically the Born approximation (that the environment is weakly perturbed by the system), and the Markov approximation (that the environment memory time is short compared to the system dynamics).
Such a breakdown  is the most commonly stated motivation for a non-Markovian treatment.
There are many ways in which these approximations may fail.  These include:
\begin{itemize}
    \item When the coupling to the environment is strong. Here the approximations could break down either because the resulting dissipation becomes fast, so that timescale separation fails, or because the Born approximation fails.    
    
    \item When the environment is highly structured, with sharp peaks.  In the time domain this corresponds to resonant vibrations which oscillate for a long time before decaying.
    There are several ways this might arise.  It can arise because of sharp resonant modes in the environment.  This is relevant when, for example, modeling the effect of vibrational modes in organic molecules or polymers.  Structured environments can also occur due to interference effects, so that coupling to modes at particular wavelengths might be suppressed.  This can occur for problems involving electronic excitations moving between different sites coupled to delocalised environment modes, such as for a double quantum dot coupled to phonons~\cite{Hall2025Controlling}.

    \item When the environment is dominated by slow modes so that memory times become long.   
    This can arise in any system where there is a large density of states of low frequency modes.  This can be something that occurs intrinsically due to the structure of the system (such as fermionic problems as mentioned below).   
    It can also be something which occurs when the system is tuned close to a continuous phase transition.
    In such cases, the critical slowing-down that occurs close to a phase transition leads to a large density of states for slow modes, so that Markovian descriptions may fail~\cite{Nagy}.

\end{itemize}

Having discussed some key ``types'' of non-Markovian systems, we may also now mention some specific examples.
One extensively studied context where these issues arise is light absorption in biological light-harvesting systems~\cite{scholes2011lessons,huelga2013vibrations,lambert2013quantum,wang2019quantum,Cao2020Quantum}; here it has been established that vibrational resonances can be crucially important in determining how energy is transferred. 
Light harvesting complexes (in line with many organic chromophores) typically have complex vibrational spectra, associated with the different stretching and bending modes of the molecule.  This leads to a peaked spectral density, for which Markovian approximations fail.

More broadly, non-Markovian situations arise across a range of fields, including in chemical physics, e.g. electron transfer dynamics in molecules and in solution, and in condensed matter physics, e.g. in the form of impurity problems.  In the latter case, prominent examples are models such as the Kondo or Anderson models~\cite{hewson1997kondo} describing impurity spins coupled to fermionic environments, or the spin-boson model~\cite{Leggett1987dynamics} which may arise when bosonizing such a fermionic environment or when a small quantum system interacts with phonons in a solid.  In these cases, the coupling to the environment can be strong.
Moreover, the environment can generally be expected to have a high density of states of slow, low energy modes.  Such behavior will typically arise in fermionic problems with a Fermi surface, where there are naturally a continuum of low frequency bosonic modes associated with deformations of that surface.
As discussed further below, these paradigmatic models often serve as benchmarks for non-Markovian methods.

In addition to the examples discussed above, there can be cases where the complexity of the system forces one to consider non-Markovian treatments.   The standard Markovian approaches (e.g. Bloch--Redfield derivation of the master equation) require that one can determine the dynamics of the system in order to find the response of the environment to that dynamics.  For complex systems, this may be challenging.  Indeed, it is this issue that has led to debates about the use of ``local'' vs ``global'' master equations---i.e.\ whether one can get away with using an approximate form of the system dynamics to derive the master equation~\cite{gonzalez2017testing,hofer2017markovian,cattaneo2019local}. One context where this question arises is in models of molecules strongly coupled to light and forming polaritons. Here, strong light-matter coupling modifies the dynamics of the molecules, so that it is only possible to perform the derivation of the master equation in limiting cases, such as in the single-excitation subspace.
Indeed, any system Hamiltonian describing an interacting many-body problem can pose a similar problem.
Notably, non-Markovian treatments do not pose this same dilemma, as the environment response is instead found from the actual system dynamics.   Of course, sufficiently complicated systems may still be hard to simulate, but if the dynamics can be simulated, it is the response to that actual dynamics that is used.

\paragraph{The question does not permit it.} 
The second category of why Markovian approaches fail arises most naturally when directly asking about information return from the environment, or communication between different systems through a shared environment.  By definition, these questions ask about non-Markovian effects.
In addition, recent exploration of the effects of strong coupling to the environment on thermodynamics~\cite{talkner2020colloquium} necessarily requires one to consider effects of strong system-environment coupling and so must extend beyond Markovian approaches.

A more subtle case involves scenarios where one needs to calculate multi-time correlations.  
It is well known that even two-time correlations suffer from inevitable issues in Markovian treatments~\cite{Ford1996ThereIsNo}.
Fundamentally this occurs because Markovian treatments involve pre-evaluating the response of the environment to the system at a specific set of system frequencies.  
Multi-time correlations can however probe the response at arbitrary frequencies, extracting information which is fundamentally not present in the Markovian treatment.
In addition,  multi-time correlations can directly probe how entanglement with the environment can be created and later recovered~\cite{Milz2019Completely}, again a fundamentally non-Markovian feature.  

Multi-time correlations are physically relevant when considering spectroscopic measurements.
Calculating an optical absorption or emission spectrum of a material fundamentally requires calculating two-time correlations of the electric dipole moment~\cite{Scully1997:qo}, and then Fourier transforming the result.  
As such, any scenario where spectra are measured can yield a case where Markovian treatments give incorrect results;  for example, failing to satisfy relations required between absorption and emission spectra in thermal equilibrium~\cite{Ford1996ThereIsNo}.
Furthermore, there is a growing field of multidimensional spectroscopy~\cite{Mukamel,cho:2008,collini:2021}, where experiments use carefully designed pulse sequences with multiple pulses to measure nonlinear response of the system.
Such experiments explicitly involve questions of how operations with different time delays interact, and can thus directly probe non-Markovian effects.  We will discuss such experiments in some detail in Sec.~\ref{sec:Future:Apps:Spectroscopy}.

\subsection{Dimensional reduction and tensor networks}

In many areas of computational physics---particularly those modeling quantum systems---a key challenge is the large size of state space, required due to the exponential dependence of Hilbert-space dimension upon the number of degrees of freedom. 
To address such challenges and make large-scale computations practical, one requires approaches that allow one to effectively compress the representation of such large spaces.  

A powerful approach to such compression involves identifying the structure of the states one is trying to describe, and choosing a representation that mirrors that structure so as to make clear which correlations exist and which may be neglected.
This then allows adaptive routines where one can tune some parameter or parameters controlling the scale of correlations that are kept, and thereby trade computing time for accuracy.
In the ideal case, this trade-off can permit one to reach the desirable levels of accuracy while remaining within acceptable computational costs.
This enables one to tackle otherwise computationally complex problems, without requiring approximations to be made ``by hand.''  

For the problems we are interested in, one key way to form such a representation is through tensor networks~\cite{Verstraete2008matrix, Schollwoeck2011DensityMatrix, Orus2014Practical,Bridgeman_2017, Silvi2019tensor,Ran2020tncontr,Cirac2021rmp,Okunishi2022,Banuls2023Tensor}.    
In a tensor network, a high-order tensor is written as a network of low-order tensors, hopefully with restricted dimensionality of each  index. 
In the context of quantum many-body systems, 
such an approach is most familiar as a representation of the quantum wavefunction, particularly in one dimension, where matrix product states (MPS) are constructed based on the restricted extent of correlations between different sites on a lattice~\cite{PerezGarcia2007,Verstraete2008matrix,Schollwoeck2011DensityMatrix}.
Here, a matrix is associated with every basis state of each lattice site, such that a given configuration of the sites leads to a sequence of matrices, and the amplitude of the overall quantum state is found by the product of these matrices. 
The amount of correlations between different sites is determined by the dimensionality of these matrices---known as the bond dimension.
Broadly speaking, MPS with finite bond dimension can approximate states with low entanglement (see~\cite{Schuch2008} for rigorous results), so that in such cases the amount of information required to describe the quantum state need not grow exponentially with the number of sites.
Using this representation one can then efficiently perform time evolution of the quantum state (as long as the entanglement remains bounded), or perform variational minimization to find the ground state.

\begin{figure}
    \centering
    \includegraphics[width=\linewidth]{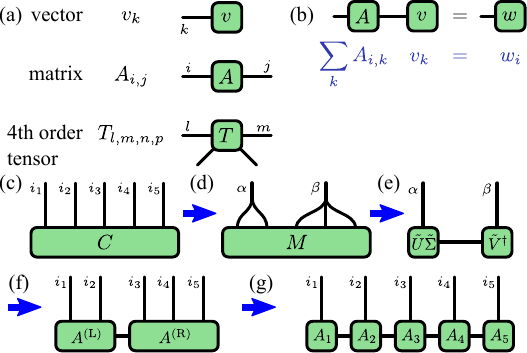}
    \caption{Penrose graphical notation for tensor networks.
    (a) Vectors, matrices, and tensors are represented as shapes, with each ``leg'' corresponding to an index.
    (b) A leg connecting two tensors indicates summation over the ``contracted'' index.
    (c-g) Matrix product state decomposition of a fifth-order tensor.  (c) Shows the original tensor.  By grouping indices (d) puts this in the form of a matrix, to which a singular-value decomposition can be applied (e).  After reshaping to the original indices this yields (f).  Repeated application decomposes the high order matrix (c) into a network of low order matrices (g). }
    \label{fig:tensor-network-notation}
\end{figure}

Tensor network literature makes extensive use of a diagramatic approach based on Penrose's graphical notation~\cite{Penrose1971Applications}.
While this perspective does not aim to be a complete pedagogical introduction to tensor networks, Fig.~\ref{fig:tensor-network-notation}(a,b) provides a brief summary of that notation which we use throughout this perspective.
In addition Fig.~\ref{fig:tensor-network-notation}(c-g) provides an illustrative example of how a high-order tensor (c) could be decomposed into a network of lower order tensors, specifically a matrix product state as shown in Fig.~\ref{fig:tensor-network-notation}(g).

Matrix product state (MPS) approaches have been generalized in several ways, including tensor network states in higher dimensions~\cite{Verstraete2004b} or with different connectivity~\cite{Shi2006ttn,Vidal2008mera}, describing density matrices rather than wavefunctions~\cite{Verstraete2004a,Zwolak2004,Pirvu2010mpo}, 
and describing problems in the continuum limit (i.e.\ without a lattice) using continuous tensor network ansatzes
~\cite{verstraete2010continuous,Haegeman2013calculus,Haegeman2013cmera,Tilloy2019cpeps}.
The same concept of tensor decomposition has been known in other fields; in applied mathematics and computer science, such decompositions have been known as tensor trains~\cite{Oseledets2011TT}, and in classical statistical physics, where they were introduced for contracting partition functions variationally,~\cite{Baxter68,Okunishi2022} and they have been known as a route to find \emph{exact} solutions for classical transport problems~\cite{Derrida1993Exact}.
Recently, there have been a number of new applications of such methods in different contexts.
These have included using matrix product states to represent continuous functions, and thus the time evolution of partial differential equations~\cite{GarciaRipoll2021quantuminspired,gourianov2022quantum,ye2022quantum}, and the use of ``tensor-train cross interpolation'' to use matrix product states to replace Monte Carlo integration over high dimensional integrals~\cite{NunezFernandez2022Learning}.

\subsection{Outline} 

The remainder of this perspective is organized as follows.
Section~\ref{sec:PT} will introduce the key object of our discussion,  the process tensor, and in particular its calculation as a matrix product operator in Sec.~\ref{sec:PT:computing}.  As part of this, in Sec.~\ref{sec:PT:unifying}, we will show how the process-tensor approach can be related to many other approaches to non-Markovian dynamics.  
Section~\ref{sec:benchmarking} discusses questions regarding benchmarking different algorithms and their numerical implementations, including comparisons to methods not involving process tensors.
Section~\ref{sec:Future:Apps}  discusses classes of problem for which process tensor approaches are particularly suitable.
Finally, in Sec.~\ref{sec:future}, we will discuss possible future research directions, both in terms of methods which may extend the range of systems that can be modeled, and in terms of application to questions that can be addressed with these methods.

\section{The Process Tensor formalism}
\label{sec:PT}

As motivated above, this section introduces the mathematical formulation of the \emph{process tensor}.
It is both a convenient  abstraction for describing arbitrary dynamics of non-Markovian open quantum systems as well as a practical computational tool. 
The latter is because it admits an efficient tensor network representation in many physically relevant cases, allowing efficient computations of non-Markovian dynamics.
The process tensor~\cite{Chiribella2008,Chiribella2009,Pollock2018Operational,Pollock2018NonMarkovian} is a multilinear map from a set of operations on a quantum system to its final state.
Those operations can include simple ones like evolution under a (time-dependent) Hamiltonian, but can also include processes such as state preparation, measurement, or action by an arbitrary quantum gate.
Clearly, knowing such an object is very powerful, however one might suspect that it will be  exponentially complex to calculate and store.
Remarkably, this is not the case for many relevant scenarios, and efficient tensor network representations of the process tensor can exist.
In the following we will provide a brief summary of the process tensor, and highlight how it can provide a unified understanding of different approaches to non-Markovian dynamics.

\subsection{Brief Introduction to Process Tensors}
\label{sec:PT:intro}

This section aims to briefly introduce the concepts and technical manipulations involved in treating non-Markovian dynamics with tensor networks, focusing on the process tensor formalism at a high level. For more details on the technical steps and definitions, the presentation below is based on Refs.~\cite{Link2024Open,cygorek2024sublinear,Cygorek2022SImulation,Jorgensen2019Exploiting}. 
 
\begin{figure}[ht]
    \centering
    \includegraphics[width=0.7\columnwidth]{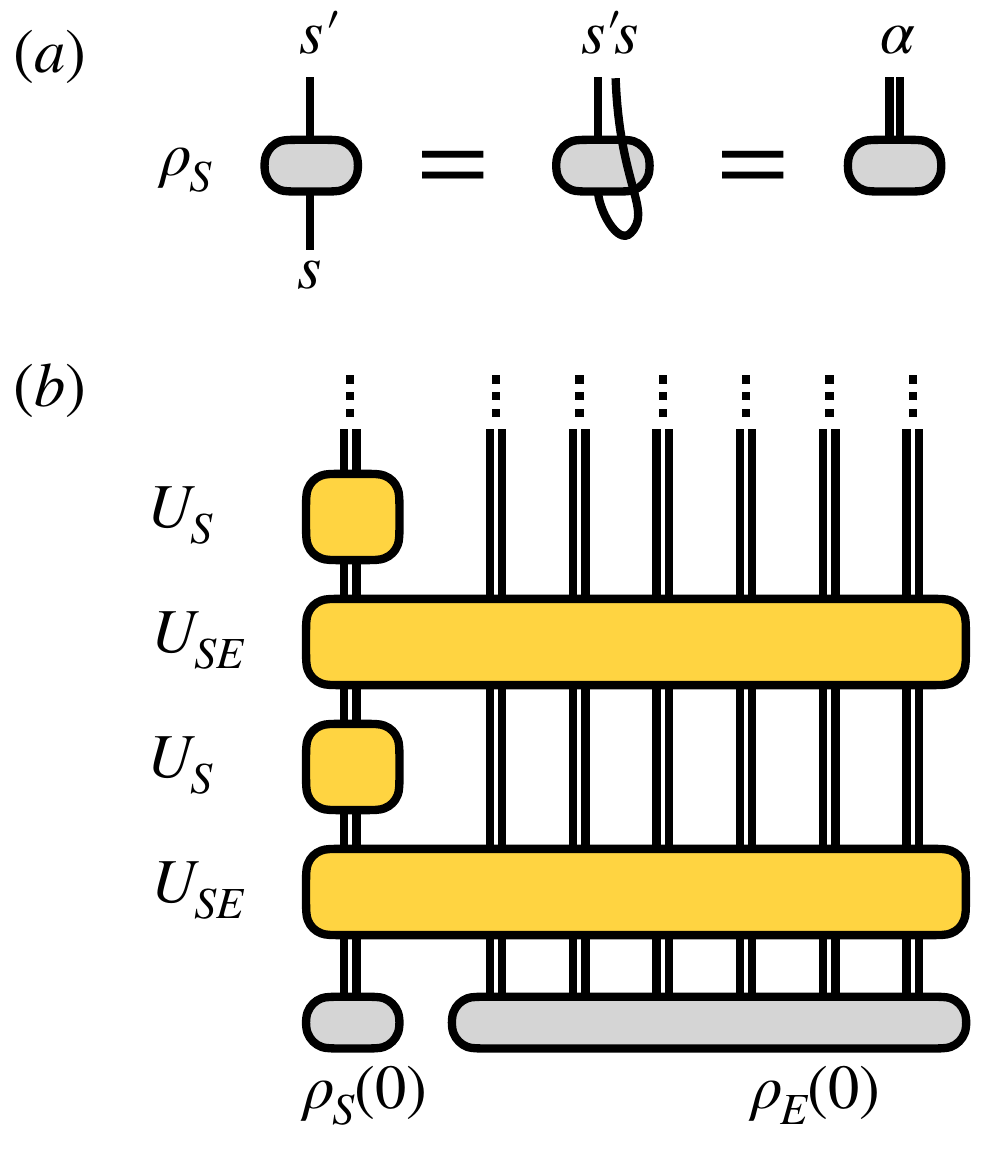}
    \caption{ 
    \label{fig:Liouville_formalism} (a) Mapping of system density matrix to single Liouville-space index $\alpha=(s',s)$ and (b) evolution of the combined system (S) and environment (E) density matrix, which is taken to be a product at time $t=0$, and evolved by split or Trotterized time-evolution operators $U_S$, describing the system evolution only, and $U_{SE}$, which captures the system-environment interaction and environment dynamics.
    In this graphical language, each shape can be identified with a tensor and its legs indicate  the indices of that tensor (e.g.\ $\rho_S$ in panel (a) is a matrix or order-2 tensor). A leg connecting two tensors represents a contraction over the corresponding index. When legs are explicitly labeled, the diagram represents a particular component of the tensor (e.g.\ in panel (a)), otherwise the diagram represents the full tensor. A tensor network, constructed from contractions of individual tensors, has the dimension corresponding to its open legs (e.g. panel (b) corresponds to the full density matrix tensor after two time steps). 
 }    
\end{figure}

In the context of open systems, the state of the system $S$ is defined by a density matrix $\rho_S$, which can be expressed in a basis of states $\ket{s}$ where the
$s=1,2,...d$ is an index that enumerates the states in a Hilbert space of dimension $d$. Then the system density matrix is the matrix $\rho^{s' s}_{S} = \bra{s'} \rho_S \ket{s}$.
For convenience, it is common practice in the open system literature to combine the $s'$ and $s$ indices into a single \emph{Liouville-space index} 
$\alpha = (s',s)$ that runs from 1 to $d^2$.
Figure~\ref{fig:Liouville_formalism}(a) shows this folding of indices into the doubled Liouville index $\alpha$.
Practically speaking, this mapping folds the ``bra'' and ``ket'' indices together so that the resulting formalism resembles the evolution of pure states, while actually describing mixed states. Another way to describe this process is that one starts with density matrices and superoperators (linear maps on density matrices) and ``vectorizes'' these objects into state vectors and operator matrices, respectively.

One is interested in the dynamics of the system $S$ coupled to a large or infinite environment $E$ (also often called the ``bath''). The combined system and environment (SE) is described by a density operator $\rho(t)$ that is evolved in time by a unitary operator, \mbox{$U(t+\Delta,t) \rho(t) = \rho(t+\Delta)$}. (Note that the system density matrix is related to $\rho(t)$ by tracing over the environment to obtain  $\rho_S(t) = \text{Tr}_E[\rho(t)]$.) When the time step $\Delta$ is small, the time-evolution operator $U$ can be approximately factorized via the Trotter formula as a product of two separate
evolution operators $U(t+\Delta,t) \approx U_S(t+\Delta,t) U_{SE}(t+\Delta,t)$. The operator $U_{SE}$ captures the coupling of the system with the environment as well as the dynamics the environment itself, whereas $U_S$ only includes terms acting on the system. In this form, the entire evolution can be described, without loss of generality, in the form of the layered tensor network as shown in Fig.~\ref{fig:Liouville_formalism}(b).

\begin{figure}[t]
    \centering
    \includegraphics[width=1.0\columnwidth]{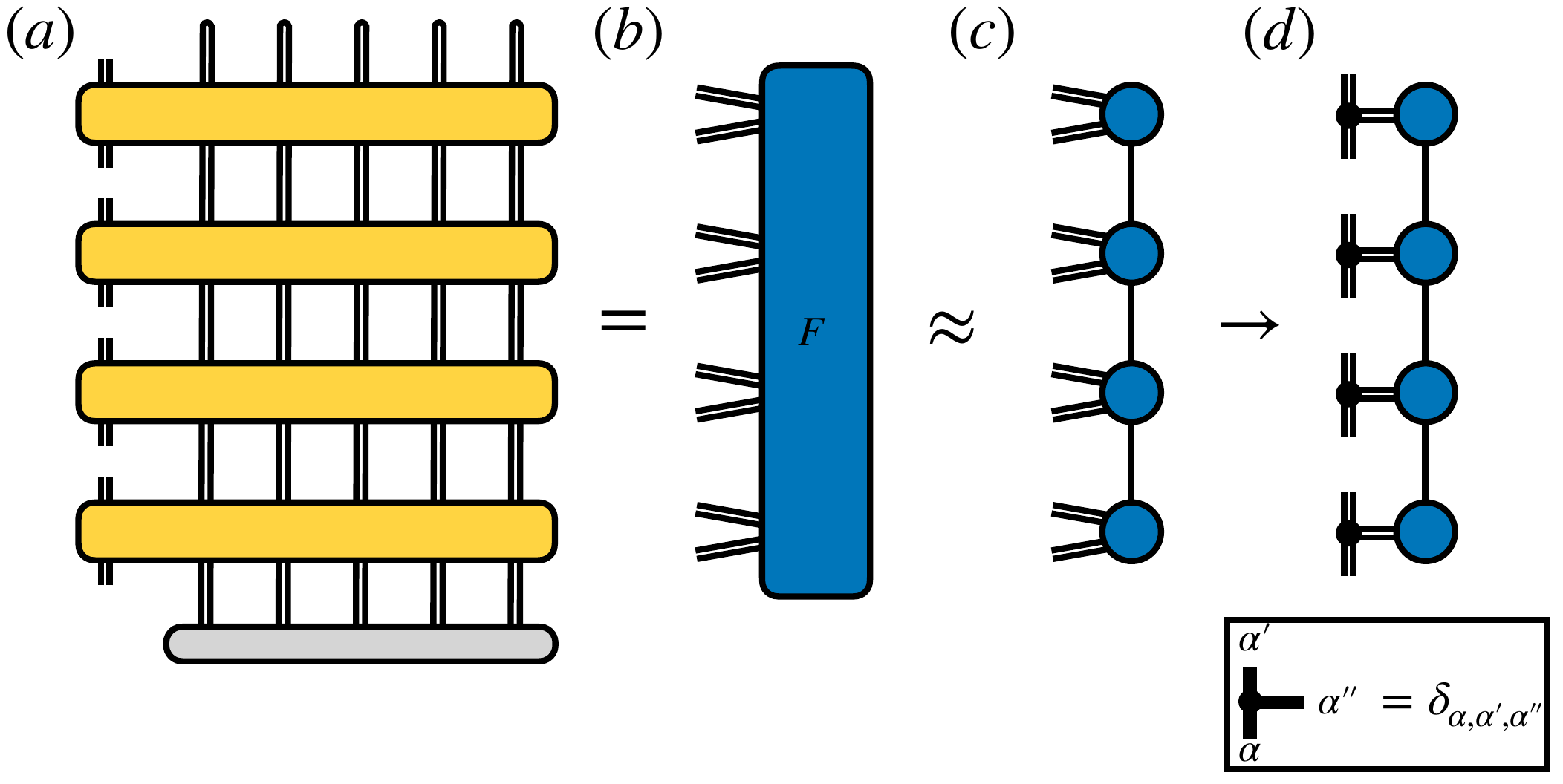}
    \caption{ 
    \label{fig:process_tensor} 
    Definition of the process tensor and approximation as a tensor network. 
    Panel (a) shows the time evolution network of Fig.~\ref{fig:Liouville_formalism} with the environment traced out and the system evolution operators $U_S$ removed.
    The resulting tensor in panel (b)  is the process tensor $F$.
    The environment is assumed to initially be factorized from the system, but this assumption can be relaxed. In panel (c), $F$ is approximated by a matrix product operator tensor network.
    In some cases we can assume the system-environment coupling can be chosen  diagonal in the system basis.  In this case, as shown in panel (d), the process tensor can be rewritten as a matrix product state, along with a generalized Kronecker $\delta$ tensor (also called a copy tensor) as defined in the box.  }
\end{figure}

The next step in the formalism is to define a network of tensors where the environment degrees of freedom have been traced out and with the system evolution operators $U_S$ removed (thus only the combined system-environment operators $U_{SE}$ remaining), as shown in more detail in Fig.~\ref{fig:process_tensor}(a). 
The resulting network defines the so-called process tensor $F^{\alpha_1\alpha_1^\prime \alpha_2 \alpha_2^\prime \alpha_3 \alpha_3^\prime \cdots \alpha_N}$ where the
subscripts on the $\alpha_j$ indices refer to the discretized time steps of size $\Delta$, with a total of $N$ time steps.
The two indices $\alpha_n, \alpha_n^\prime$ correspond to the ``input'' and ``output'' states to $U_{SE}$ at timestep $n$.
In many relevant cases (see below), one may work in a diagonal basis of the system environment coupling, in which case these two indices $\alpha_n, \alpha_n^\prime$ are constrained to be equal.  In such cases the process tensor simplifies to exactly $N$ indices, $F^{\alpha_1 \alpha_2  \alpha_3  \cdots \alpha_N}$.

Because the process tensor has a number of indices growing linearly with the number of time steps $N$, representing it naively as a dense tensor would be exponentially costly in the number of time steps. 
Fortunately, for many environments of interest the process tensor admits an efficient representation as the tensor network as depicted in Fig.~\ref{fig:process_tensor}(c,d).
When we allow the general case, where the environment can change the system state, this corresponds to a network where each time step has a four-index tensor, with two indices into the system Hilbert space (input and output indices), and each tensor
connected to other tensors in the network by a bond index of some moderate size $\chi$. 
This case can naturally be thought of as a matrix product operator (MPO) representation of the process tensor (i.e.\ a PT-MPO).
In the special case where the environment coupling is diagonal, then one can simplify further with each time step captured by a  three-index tensor.
Here one could consider this as a matrix product state (MPS) representation of the tensor.
To retain generality in subsequent discussions, we will refer to the MPO representation unless discussing statements that apply only to the MPS form.
The tensor network decomposition reflects the locality of correlations in time and the intuition that system-environment interactions at a time $t$ have a diminishing effect on the system at later times $t' > t$ for large enough values of $|t'-t|$.

One way to understand the role of the tensor network decomposition of the process tensor is by analogy to classical non-Markovian systems.
A widespread approach to modeling such systems is by identifying a ``Hidden Markov Model''~\cite{rabiner1986introduction}.  This means expanding the system state space sufficiently to include extra hidden degrees of freedom such that the resulting dynamics becomes Markovian.  
There are subtleties about extending such ideas to quantum systems~\cite{Milz2021Quantum,Taranto2025HigherOrder} (particularly around the failure of the Kolmogorov extension theorem~\cite{Milz2020kolmogorovextension}).
However, one can view the tensor network decomposition of the process tensor as defining an effective set of hidden quantum degrees of freedom such that the resulting evolution of system and hidden degrees of freedom becomes Markovian.
This idea was used in Ref.~\cite{Pollock2018Operational} to provide an operational definition of non-Markovianity in terms of process tensors.

In terms of practical calculations, it is important to note how the process tensor isolates the evolution of the bath from the possible interventions on the system~~\cite{Chiribella2008,Pollock2018Operational}, which can then be varied. Thus, even when the process tensor approach is not advantageous for a single calculation, there may still be benefits for problems where the process tensor is used many times once calculated---see Sec.~\ref{sec:Future:Apps}.

Many rather different methods exist by now to compute the process tensor directly in its tensor network form, bypassing the exponential cost of representing it fully.
These methods range from boundary contraction of path integral circuits~\cite{Banuls2009}, tensor network contraction~\cite{Jorgensen2019Exploiting,Cygorek2022SImulation}, decompositions of Slater determinants into MPS~\cite{Thoeniss2023Efficient, Ng2023Realtime, Park2024tensor}, and algorithms for applying tensors to infinite MPS~\cite{Link2024Open}. 
To make the derivation clear, we present in detail two examples for two different types of environment:  We first present the infinite MPS method for bosonic baths, since it is one of the most straightforward derivations.  We then summarize the Fishman--White algorithm for a fermionic environment.
In the sections following, we review several further approaches that can alternately be used, starting from other methods for non-Markovian open quantum systems.  These sections illustrate how the process tensor can unify many approaches to non-Markovian dynamics.

\subsection{Computing the Process Tensor of a Free Environment}
\label{sec:PT:computing}

\subsubsection{Free Bosonic Environment}

From the previous section, it may seem
that computing a process tensor and converting it to an MPO tensor network (Fig.~\ref{fig:process_tensor}(c))
would be prohibitively expensive. However, in many practically relevant situations for bosonic environments, the process tensor may already be given as a tensor network,
such as a circuit of time evolution gates. Then the task of computing the process tensor MPO reduces to the task of converting
a circuit into an MPO via contraction and compression.

To make the conversion of a process tensor into an MPO  more concrete, we will review a common case in which
the process MPO is computable using a standard tensor network algorithm known as the iTEBD algorithm. 
This section is based on the approach of Link, Tu, and Strunz~\cite{Link2024Open}. Additional background on
the quantities involved can be found in Refs.~\cite{cygorek2024sublinear,Jorgensen2019Exploiting}.

In this section, we consider a  Hamiltonian of the form
\begin{align}
\hat{H} & = \hat{H}_S + \hat{H}_{SE} \nonumber \\
\hat{H}_{SE} & = \hat{O}_S \, \sum_k (g_k \hat{a}^\dagger_k + g^{*}_k \hat{a}_k) + \sum_k \omega_k \hat{a}^\dagger_k \hat{a}_k 
\label{eq:bosonic-SE-coupling}
\end{align}
where the system Hamiltonian $H_S$ can be general and we consider an arbitrary Hermitian system-environment coupling 
operator $\hat{O}_S$. 
The environment is a system of free bosons $\hat{a}_k$ with single-particle energies $\omega_k$
and Hamiltonian $\hat{H}_E = \sum_k \omega_k \hat{a}^\dagger_k \hat{a}_k$.
One can assume the system basis is chosen to make the coupling operator
diagonal: 
$\hat{O}_S = \sum_s \lambda_s \ket{s}\bra{s}$. 

We note that while there is no loss of generality in assuming the Hermitian operator $\hat{O}_S$ can be written in the basis in which it is diagonal, there is a loss of generality in assuming the form of system-environment coupling in Eq.~\eqref{eq:bosonic-SE-coupling}.  
This restriction is in fact the special case, mentioned in the previous section, for which the process tensor can be reduced to an MPS rather than requiring an MPO, and it is this case which we discuss in the rest of this section.
The most general coupling would not necessarily have the form of a single product of a system operator and an environment operator,  but would instead be a sum of several such terms.  An alternate approach to deriving process tensors MPOs for this general case has been proposed in Ref.~\cite{Richter2022Enhanced}, at the expense of extra indices in the tensor network.

Because of the Gaussian form of the environment, one can derive an exact expression for the process tensor~\cite{Jorgensen2019Exploiting}, namely
\begin{align}
F^{\alpha_1 \alpha_2 \cdots \alpha_n} & = \prod_{1 \leq j \leq i \leq n} b_{i-j}(\alpha_i, \alpha_j) \\ \nonumber
& = \prod_{\substack{\ell=0,\ldots,(n-j) \\j=1,\ldots,n}}  b_\ell(\alpha_{\ell+j},\alpha_j),  \label{eq:bosonic_process_tensor}
\end{align}
where the integer indices $i,j=1\,\ldots\, n$ label the time steps, and
the factors are given by
\begin{align}
b_{i-j}(\alpha_i, \alpha_j) & = b_{i-j}(s_i, r_i, s_j, r_j) \\ \nonumber
& = \exp\left[ - (\lambda_{s_i} - \lambda_{r_i}) (\eta_{i-j} \lambda_{s_j} - \eta^{*}_{i-j} \lambda_{r_j})\right] \ .
\end{align}
The $\lambda_{s}$ are the matrix elements of the operator $\hat{O}_S$ in the system basis, and the factors $\eta_{i-j}$
are given by integrals of the environment correlation function across pairs of time windows between the time points at which the 
process tensor is evaluated. (See Appendix~\ref{app:details} for explicit expressions for the $\eta_{i-j}$ factors and definition of the environment correlation function.) 
Thus the $\eta_{i-j}$ factors capture the effect of the environment on the system, which is non-local but decaying in time.
Asymptotically, the $\eta_{i-j}$ decay as $\eta_{i-j} \propto C(\Delta t \cdot |i-j|)$ where $C(t)$ is the environment correlation function
and $\Delta t$ is the time step used for discretization of the system evolution.

\begin{figure}[t]
    \centering
    \includegraphics[width=0.85\columnwidth]{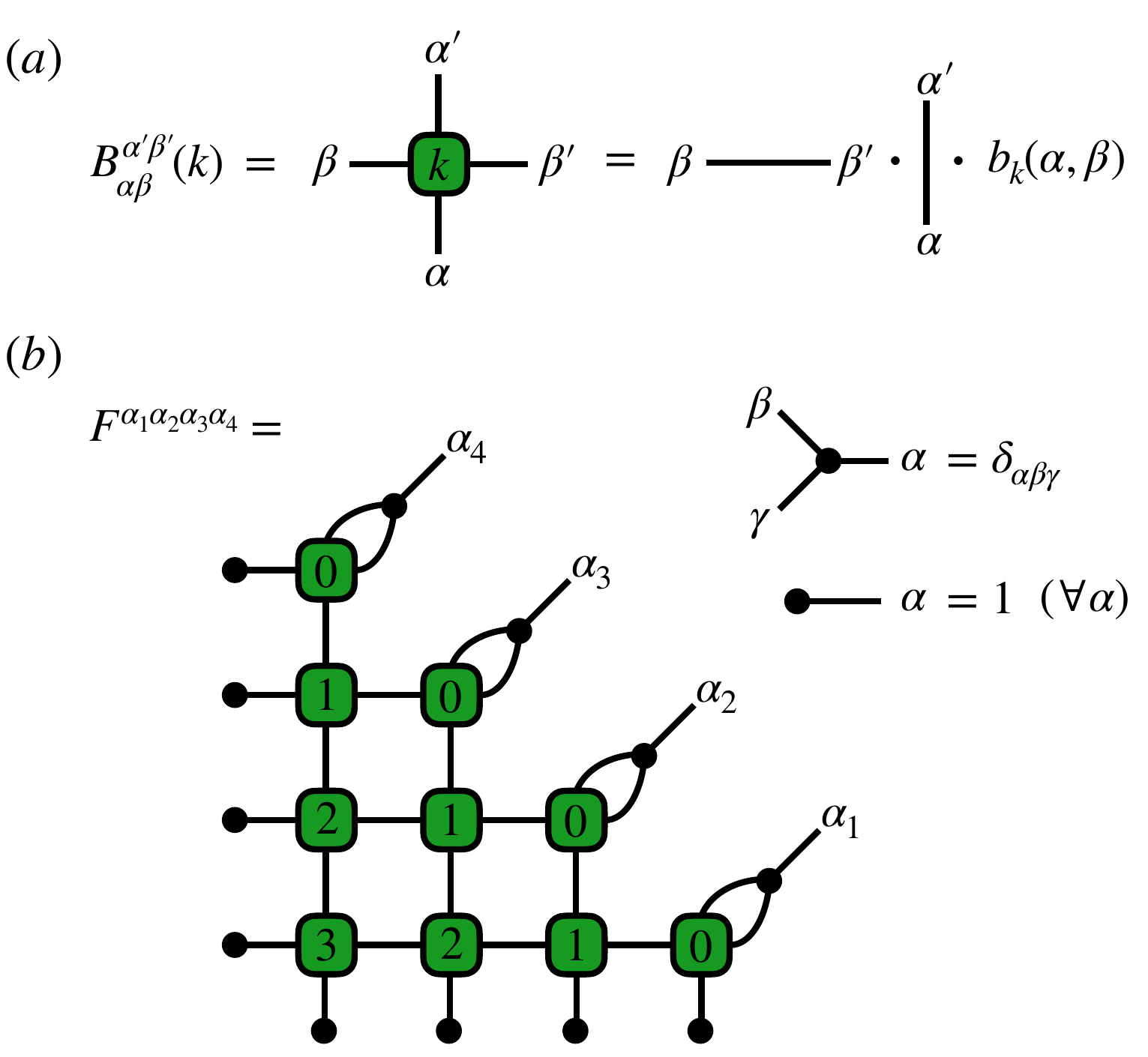}
    \caption{ 
    \label{fig:boson_F_network}
    Defining the tensor $B$ as shown in panel (a) leads to the tensor network expression (b) for the process tensor of a
    Gaussian boson environment (shown here for the case of four time  indices $\alpha_1$--$\alpha_4$).
    }
\end{figure}

The explicit form of the process tensor above can be brought into the form of a tensor network 
by defining the tensor $B^{\alpha' \beta'}_{\alpha \beta}(k) = \delta^{\alpha'}_{\alpha} \delta^{\beta'}_{\beta} b_k(\alpha,\beta)$.
The expression for $F$ now takes the form shown in Fig.~\ref{fig:boson_F_network}, with different tensors in the network accounting
for correlations between different external indices of $F$. For example the tensor labeled ``3'' accounts for correlations 
three time steps apart. A common approximation is to take a finite maximum memory time $t_\text{max} = \Delta t \cdot N_m$, which means approximating $b_k(\alpha,\beta) \approx 1$
thus $B^{\alpha' \beta'}_{\alpha \beta}(k) \approx \delta^{\alpha'}_{\alpha} \delta^{\beta'}_{\beta}$ for $k > N_m$.

\begin{figure}[b]
    \centering
    \includegraphics[width=\columnwidth]{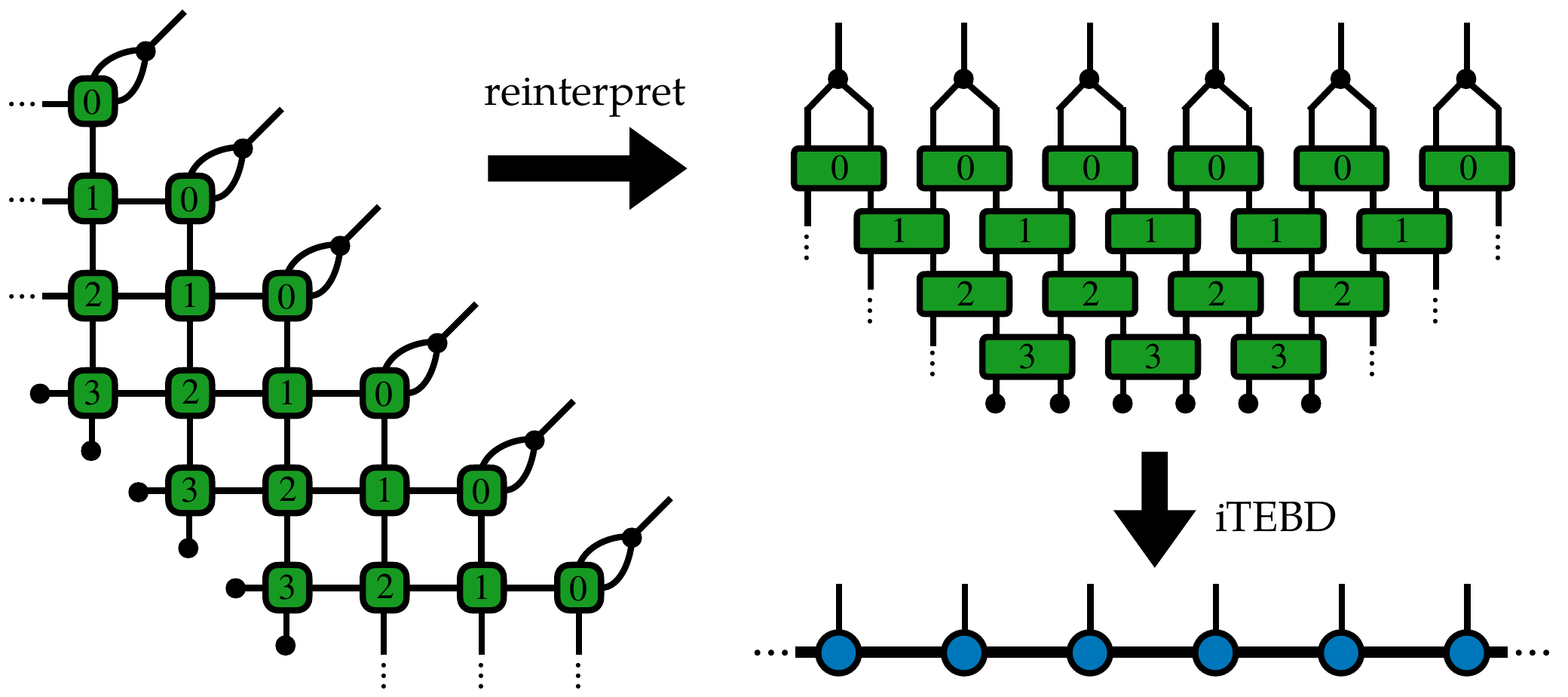}
    \caption{ 
    \label{fig:F_iTEBD}
    Reinterpretation of an infinite-time, process tensor of Gaussian bosons as a shallow quantum circuit by taking a
    finite memory time approximation. The iTEBD algorithm produces a controlled approximation of the state computed by
    such a circuit as an MPS tensor network.
    }
\end{figure}

The insight of Link et al.~\cite{Link2024Open} is that for a finite memory time and infinitely many time steps, 
the tensor network for $F$ can be interpreted as a shallow, infinitely wide quantum circuit made of non-unitary gates. 
See Fig~\ref{fig:F_iTEBD} for diagrammatic depiction of this process. Transforming an infinite circuit into an MPS tensor network can be done with the technical, yet straightforward infinite time-evolving block decimation (iTEBD) algorithm~\cite{Orus2008iTEBDBeyond} used frequently in the tensor network literature. 
The result is an approximation of the process tensor as an infinite, or time translation invariant matrix product state tensor network.

The advantages of the approach include its simplicity and its favorable scaling,
which is just $\mathcal{O}(N_m \log(N_m))$, where $N_m$ is the number of memory steps or depth of the circuit used~\cite{Link2024Open}.
The efficiency is further enhanced because the MPS bond dimensions grow slowly throughout the early layers of the circuit, 
since $B(k)$ is close to a ``swap'' operator (acting initially on a product state) for large $k$ and 
only becomes more ``entangling'' as $k\rightarrow 0$.

\begin{figure}
    \centering
    \includegraphics[width=\linewidth]{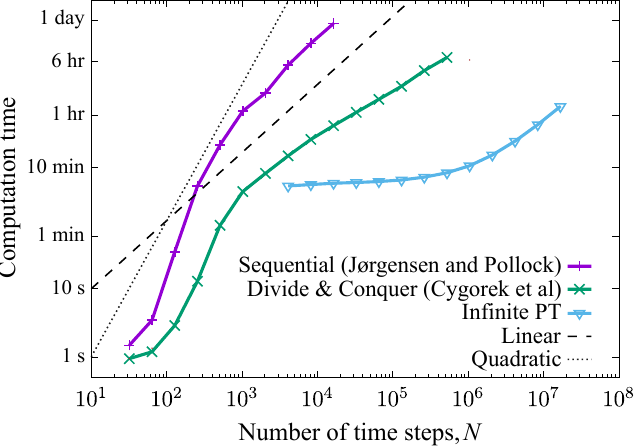}
    \caption{Comparison of computation time for different process tensor approaches to calculating the fluorescence spectrum of a quantum dot as a function of number of simulated timesteps.   
    Results are shown for the sequential algorithm Ref.~\cite{Jorgensen2019Exploiting}), the divide-and-conquer scheme (Ref.~\cite{cygorek2024sublinear}), and the time translation invariant (i.e. infinite) PT approach (Refs.~\cite{cygorek2024sublinear,Link2024Open}).  
    Gray dashed and dotted lines show linear and quadratic scaling for reference.  
    Adapted from~\cite{cygorek2024sublinear}.}
    \label{fig:CygorekCompTime}
\end{figure}

If such an approach is used to follow the time evolution of an open system for $N$ time steps, the computational cost would remain constant versus $N$ for small $N$, when the cost depends on the calculation of the process tensor.  Ultimately, at large $N$, the cost will become dominated by calculating and reporting the state at each timestep.  A figure showing this (and comparison of the cost to some other process tensor algorithms) is given in Fig.~\ref{fig:CygorekCompTime}.

\subsubsection{Free Fermionic Environment}

The previous section summarizes an efficient method for determining the process tensor for a quadratic bosonic bath.
For non-interacting fermionic baths---particularly when the state of the fermionic bath is Gaussian---there are also methods that can efficiently produce a process tensor and its associated influence functional in an analogous way.  The key point is that such objects can be computed exactly in terms of anticommuting Grassmann numbers $\xi$~\cite{grifoni}.

Anticommuting variables can be represented naturally in terms of fermionic wavefunctions by using $\xi_{i_1} \cdots \xi_{i_n} \to \hat{c}_{i_1}^{\dagger} \cdots \hat{c}_{i_n}^{\dagger} |0_{i_1}, \ldots, 0_{i_n}\rangle$ where $\hat{c}_{i_n}^\dagger$ are fermionic creation operators.
Under this mapping, the representation of the fermionic influence functional $\exp[\xi_{i} M_{ik} \xi_{k}]$ becomes equivalent to the representation of the Hartree-Fock-Bogoliubov (HFB) wavefunction $\exp[\hat{c}_{i}^{\dagger} M_{ik} \hat{c}_{k}^{\dagger}] |0, \ldots, 0\rangle$.
Several algorithms exist for the efficient approximation of such pairing states~\cite{Fishman2015, Petrica2021, Surace2022fGauss,Jin2022}; we shall expand on the approach of Fishman and White~\cite{Fishman2015, Thoenniss2023} here, although other algorithms invoke similar ideas.

The Fishman--White construction makes use of the equivalent representation of HFB states in terms of $2N$-by-$2N$ correlation matrices $C$, with submatrices defined as 
$$C_{ik} = \langle \Phi_{\text{HFB}} | \begin{pmatrix} \hat{c}^{\dagger}_i \hat{c}_k & \hat{c}^{\dagger}_i \hat{c}^{\dagger}_k \\ \hat{c}_i \hat{c}_k & \hat{c}_i \hat{c}^{\dagger}_k \end{pmatrix} | \Phi_{\text{HFB}} \rangle.$$
The construction proceeds by iteratively diagonalizing $C$, tantamount to constructing the basis of natural orbitals for $|\Phi_{\text{HFB}}\rangle$.
At the $n$th step of the iterative diagonalization, the $n$th natural orbital is constructed by a sequence of nearest-neighbor unitary transformations over $N-n-1$ sites.
The local nature of this construction works well with an MPS form for $|\Phi_{\text{HFB}}\rangle$.
Importantly, if the natural orbitals are assumed to have a bounded extent $\mathrm{min}(N-n-1, \ell)$ for some choice of $\ell$, the bipartite entanglement entropy of the resulting MPS approximation will not grow extensively with $N$. 
Thoenniss et al.~\cite{Thoenniss2023} showed this to be a valid truncation procedure---in addition to usual MPS compression techniques---since the influence functional for Gaussian fermionic baths has area law entanglement when its bath autocorrelation function decays sufficiently rapidly in time.
Thus, the Fishman--White algorithm is able to find an efficient MPS approximation to the influence functional.

One downside of algorithms based on the correlation matrix $C$ is that they obscure the limiting form of the time-translational invariance of the influence functional as the propagation time increases.
Working in the original ``temporal'' basis as na\"ively appears in the influence functional, one can construct infinite MPS~\cite{Sonner2025, Guo2024} analogously to the Gaussian bosonic case as described in the previous section.  Note that the discussion above for the fermionic influence functional is based on an MPS wave function construction.  The use of a direct MPO formulation, although more involved~\cite{parker2020}, is possible using the representation outlined above and would be useful in describing the infinite time behavior.

\section{Unifying Non-Markovian Approaches with Process Tensors}
\label{sec:PT:unifying}

The process tensor concept provides a unifying framework for different numerical approaches that have been historically devised  
to deal with non-equilibrium open quantum systems.  
As such it can act as a ``Rosetta stone'' to show how different approaches can be connected.
This unifying role has been discussed recently in the context of optimal control in Ref.~\cite{OrtegaTaberner2024Unifying}, including discussion of how process tensors relate to generalized quantum master equations, hierarchical equations of motion, stochastic methods,  and  multi-configurational time-dependent Hartree approaches.
To make this perspective self contained, we discuss these links, as well as some other potential links in the following.

\subsection{Chain mappings}
\label{sec:PT:unifying:chain}
A common strategy to address the setup of a (small) quantum system coupled to a non-interacting environment  is to map the whole system-plus-bath onto a one-dimensional problem.
Such mapping is possible when the bath Hamiltonian is non-interacting and the coupling between the system and the bath is linear in the modes of the second, such as the one shown in Eq.~\eqref{eq:bosonic-SE-coupling}. This form is often referred to as \emph{star representation}, because the system is coupled to all of the bath modes. 
The bath Hamiltonian $\sum_k \omega_k \hat{a}_k^{\dagger} \hat{a}_k$ 
can be tridiagonalized by a standard Lanczos transformation, which maps it to a tight-binding or \emph{chain representation}. 
Using as a seed for the transformation  the  linear combination of bath modes that appears in the interaction term, e.g. $\sum_k g_k^* \hat{a}_k$ for the model in Eq.~\eqref{eq:bosonic-SE-coupling}, results in a one-dimensional problem in which the quantum system  is located on one edge of the chain,
directly coupled to a single bath mode~\footnote{Notice that this one-dimensional mapping only requires the free (non-interacting) nature of the bath Hamiltonian, independent of its actual spatial dimension.}.

In the case of a continuous bath, several options exist to describe it in terms of discrete modes~\cite{deVega2015discr}. 
This is, for instance, the basis of the numerical renormalization group (NRG)~\cite{Bulla2008rmp}, which uses a logarithmic discretization of the bath energies, or the Time-Evolving Density operator with Orthogonal Polynomials Algorithm (TEDOPA) method~\cite{Prior2010chain,Chin2010orth}, in which the discrete bath modes are an average over the spectral support of the bath, weighted by the spectral function and orthogonal polynomials. 
Typically, a truncation in the number of modes is used to render a finite chain, which is sufficient to capture a numerically exact description of the whole system plus bath up to a certain time in the setups of interest.

In all these cases, an MPS ansatz can be used to describe the state of the resulting one-dimensional system, and standard TN algorithms~\cite{Verstraete2008matrix, Schollwoeck2011DensityMatrix,Paeckel2019tevol} can be applied to simulate its dynamics~\cite{Prior2010chain,Chin2010orth,deVega2015thermofield}.
As long as the entanglement in the chain remains small, this is a feasible strategy that provides access to time-dependent system and bath observables.
In particular, quasi-exact bath correlators computed with this method have been used as input to obtain memory kernels~\cite{Lyu2023TT}.

\begin{figure}[htpb]
    \centering
    \includegraphics[width=\linewidth]{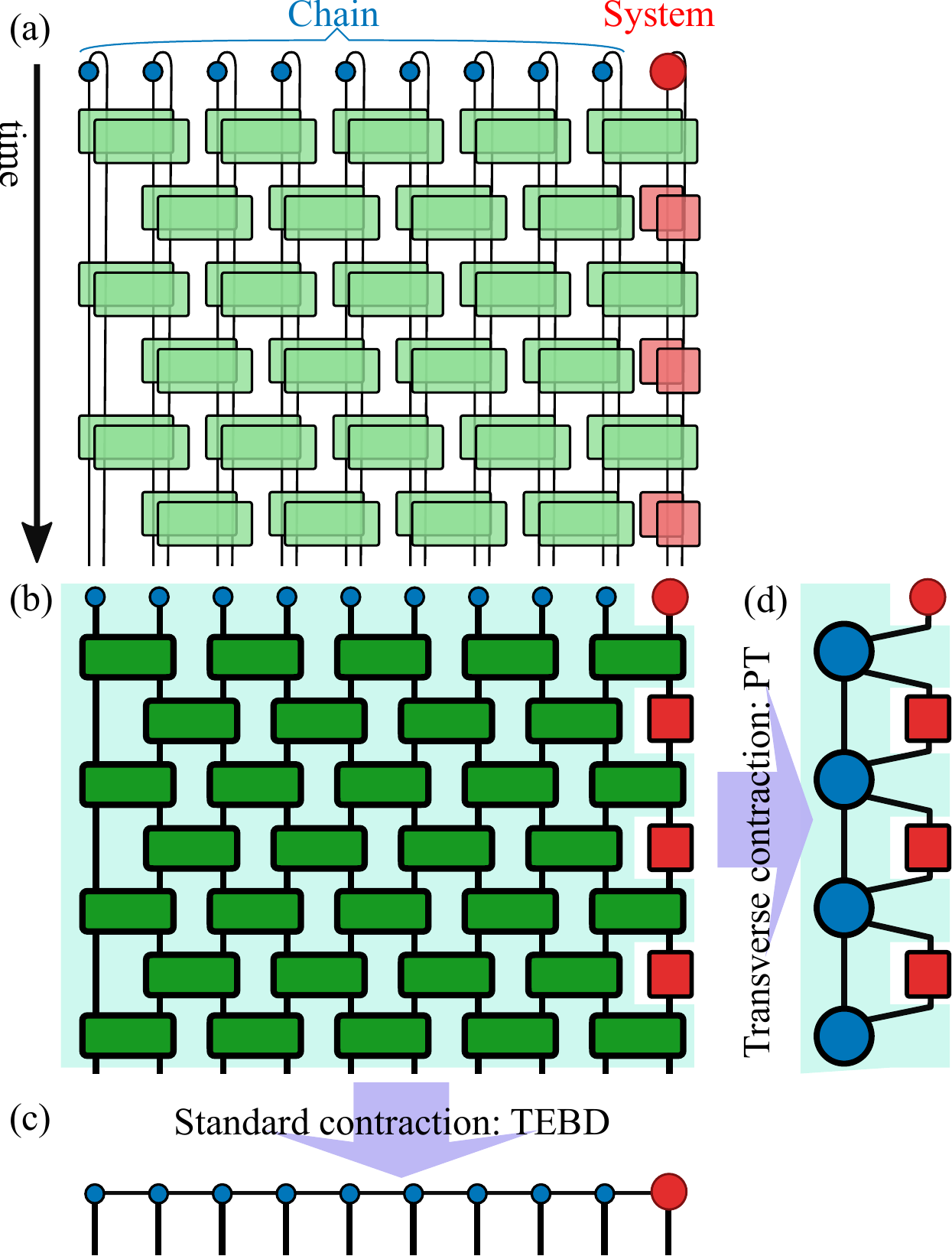}
    \caption{Chain mappings and transverse contraction. Panel (a) illustrates the application of the standard TEBD algorithm to time evolution of a chain (first 9 blue sites) coupled to a single-site system (red).  In order to evolve the density matrix, there is a double tensor network, describing the action of unitary propagation to both the left and the right of the density matrix.      
    In this figure the effects of the system Hamiltonian are taken to be described by the red gates, while the system-environment coupling and environment evolution are both contained in the green gates.
    The initial state is assumed to be a factorised density matrix of each site.
    (b) Shows the equivalent circuit after combining the two layers of the doubled tensor network to produce a tensor network in Liouville space.
    If the standard TEBD contraction is applied the result is as shown in (b):  an MPS of the chain and system describing the state after some number of timesteps.  Alternatively, if transverse contraction were applied, as shown in (c), this can produce a process tensor in MPO form (light turquoise tensors) that can then be contracted with the system evolution (red tensors).    
    }
    \label{fig:chain-tebd}
\end{figure}

But, instead of evolving the MPS ansatz explicitly, it is possible to construct a tensor network representation of the evolved system, as shown in Fig.~\ref{fig:chain-tebd}(a-b), which naturally connects this representation to the process tensor.
The unitary evolution of the whole system-plus-bath chain can be discretized using a Trotter expansion in steps of length $\Delta$, as
\begin{equation}
U(t=M \Delta)\approx \left( e^{-iH_S \Delta} e^{-i(H_{SB}+H_B) \Delta} \right)^{M}.
\end{equation}
The factor $e^{-iH_S \Delta}$ acts only on the system, corresponding to its free evolution.
Because in the chain basis the interaction and bath Hamiltonians contain only nearest-neighbour terms, the remaining factor can be further approximated as a product of two-site gates $e^{-i(H_{SB}+H_B) \Delta} \approx \prod_{j} e^{-ih_{j j+1}\Delta}$.
Each Trotter step has thus an MPO representation~\cite{Pirvu2010mpo}. 
The standard MPS evolution algorithms~\cite{Paeckel2019tevol} proceed by repeatedly applying such steps on the state and compressing the result into a MPS with limited bond dimension, but if this compression is not applied, one can also represent the evolved state as a so-called concatenated TN state~\cite{Huebener2010}, where time extends as a second dimension. 
The vectorized density operator of the whole system can be represented by a \emph{double} TN, i.e. the tensor product of the state and its complex conjugate, as shown in Fig.~\ref{fig:chain-tebd}(a).
After combining Hilbert-space indices into Liouville-space indices, this gives a network of the form shown in Fig.~\ref{fig:chain-tebd}(b).
This corresponds to a particular realization of the construction in Fig.~\ref{fig:Liouville_formalism} in which the evolution also has an MPO structure.
Tracing out the bath degrees of freedom corresponds to contracting the corresponding physical indices on both \emph{layers} for all the sites in the chain except the system one.
This results in a representation of the reduced density matrix for the system at time $t$ as a two-dimensional TN where the process tensor can be identified, as described in Sec.~\ref{sec:PT:intro}, as the part of the network that remains after cutting out the tensors that correspond to the free evolution and state of the system (shown in red in Fig.~\ref{fig:chain-tebd}).
This is an order-$(2M+1)$ tensor, with indices of dimension $d_S^2$ (the squared physical dimension of the system).
The TN form is numerically exact, but in general exponentially costly to contract.
Instead, it is convenient to try to find an approximation as a MPO of length $O(M)$ by contracting the process tensor \emph{transversely}, i.e. \emph{evolving} along the space direction.
The resulting temporal vector approximates the discrete-time Feynman-Vernon influence functional~\cite{Jorgensen2019Exploiting,Sonner2021,Ye2021}.

The transverse folding method was initially proposed to address the dynamics of local observables after quenches or the presence of magnetic impurities in translationally invariant spin chains~\cite{Banuls2009}. 
This contraction strategy of the spatio-temporal tensor network that represents the full evolution of the chain connects naturally the study of local properties of the many-body system with an open systems framework, where the chain acts as environment for a reduced subsystem.
Modified transverse contraction algorithms have also been proposed~\cite{Huang2014LongTime,Hastings2015,Frias-Perez2022Lightcone,Lerose2023over,Carignano2024}.
In all cases, a deciding factor for the method to have an advantage over standard MPS evolution techniques in the simulation of real-time dynamics is whether the so-called \emph{temporal entanglement,} i.e. the entanglement of the boundary vector in the time direction, grows more slowly than the physical (spatial) entanglement of the chain~\cite{MuellerHermes2012,Giudice2022,Carignano2024} (see~\cite{cerezoro2025} for a comprehensive discussion).

This spatio-temporal TN construction naturally connects the process tensor construction to 
more general one-dimensional quantum many-body problems. 
In this case, the system is a segment of the full chain, the rest of which acts as bath.
While the transverse contraction allows finding a TN approximation for the IF for non-integrable chains, in particular cases knowledge about the structure of the \emph{bath} wave function can be used to assist in the TN approximation.
This is notably the case for quantum impurities coupled to non-interacting fermionic baths~\cite{Thoeniss2023Efficient,Ng2023Realtime}.

\subsection{Hierarchical Equations of Motion}

In the MPO parametrization of the process tensor, the bond dimension can be understood as the smallest effective dimension of the environment required to capture its effect on the evolution of the system. This connection becomes explicit when constructing the MPO form of the influence functional from the hierarchical equations of motion (HEOM)~\cite{Link2024Open}.

In the HEOM approach~\cite{Tanimura1989Time,Tanimura2020review}, 
the effect of the environment is captured by a set of auxiliary modes, corresponding to an approximate decomposition of the environment autocorrelation function in a 
sum of complex exponentials $\alpha(t)\approx \sum_{j=1}^M \alpha_j e^{\nu_j t}$.
The non-Markovian effects are encoded in a hierarchy of auxiliary density operators $\rho_{{\bf n}}$, indexed by a vector of integers ${\bf n}=(n_1,n_2,\ldots n_M)$,
where the 0-th order
$\rho_{{\bf n}=(0,0,\ldots 0)}$ corresponds to the  density operator of the physical system, and terms with a higher index $n_k$ represent higher order interactions with the $k$-th auxiliary mode, associated to the $k$-th frequency.
These auxiliary operators evolve according to a set of coupled time-local differential equations, 
\begin{equation}
    \dot{\rho}_{\bf n}(t)=-i{\mathcal L}_S \rho_{\bf n}(t)+\sum_{{\bf n},{\bf n'}} {\mathcal D}_{{\bf n},{\bf n'}}\rho_{\bf n'}(t),
    \label{eq:heom}
\end{equation}
where the only non-vanishing terms of the dissipator ${\mathcal D}_{{\bf n},{\bf n'}}$ correspond to ${\bf n'}={\bf n},\, {\bf n}\pm {\bf e}_k$ (where $({\bf e}_k)_j=\delta_{jk}$), i.e. 
the evolution can raise or lower a single component of one of the auxiliary indices at a time.
The hierarchy, which in full would be exact, is truncated in practice to a finite number of auxiliary operators.

It is possible to think of $\rho_{\bf{n}}$ as components of the state of an extended system, consisting of the physical degrees of freedom of the system plus a set of $M$ auxiliary bosonic modes $\{n_k\}$, and to 
interpret Eq.~\eqref{eq:heom} as a single dissipative evolution equation for this quantum many-body system.
Such a perspective allows the use of time-dependent TN methods to simulate the evolution of the effective system (system plus auxiliary modes)~\cite{Shi2018heom-mps,Mangaud2023survey}, but also offers a direct connection to the process tensor, as discussed in Ref.~\cite{Link2024Open}.

Repeating the procedure described above, we can Trotterize the evolution into discrete time steps,
and concatenate them to build a TN that extends in the temporal dimension. 
This time the evolution is non-unitary, so one must start directly from the TN representation of the density matrix in Liouville space.
The process tensor can then  be identified with the section of the TN that corresponds to the auxiliary modes and their interaction with the system.
If we consider all auxiliary modes in a single site, the TN reduces to the 
contraction of an MPO with a set of matrices of the same form as appearing in Fig.~\ref{fig:chain-tebd}(d).
The set of matrices describe the free propagation of the system site, while
the MPO corresponds to the process tensor.
The bond dimension of the PT-MPO will be upper bounded by the dimension of this site containing all auxiliary modes, namely the number of auxiliary density operators in the HEOM.
In~\cite{Link2024Open} time-translational invariance was exploited to express the process tensor in terms of a unique tensor, plus suitable boundaries, but the same construction could also be used for time-dependent generators.

\subsection{Hierarchy of Pure States}

A conceptually similar approach to HEOM is the hierarchy of pure states (HOPS)~\cite{Suess2014Hierarchy}, which provides a quantum trajectory approach with a structure that resembles that of the HEOM.
Starting from the Feynman-Vernon path integral, the propagator of the system density matrix is written as an ensemble of stochastic propagators acting on the initial state. 
The stochastically propagated state of the system satisfies a non-Markovian quantum state diffusion equation~\cite{Diosi1998NonMarkovian}.
In the HOPS approach, this evolution equation is transformed into a hierarchy of coupled differential equations by introducing a set of auxiliary states which, as in the HEOM approach, are labeled by a set of $M$ integer indices ${\bf n}=(n_1,n_2,\ldots n_M)$ corresponding to a decomposition of the bath correlation function in $M$ modes.

As in the HEOM method, each component of the hierarchy index $n_k$ can be thought of as the state of an additional bosonic degree of freedom, such that the hierarchy can be rewritten as a single equation for an extended system.
In order to obtain a process tensor from this approach, one would need to obtain the average over noise realizations. Recently~\cite{Grimm2024stt}, it has been shown how the effect of the average over a noise field can be captured by an influence kernel, and approximated by a MPO ansatz,
for the case of a stochastic Liouville equation. A similar strategy  could in principle provide an explicit process tensor construction for the HOPS case.

\subsection{Generalized Quantum Master Equations}

An approach that is perhaps most similar to the standard treatment of Markovian open quantum systems is that of the generalized quantum master equation (GQME). This approach focuses directly on the reduced dynamics of the system~\cite{DeVega2017Dynamics}.
The most widely-used Nakajima--Zwanzig GQME  splits the full state space into \emph{relevant} and \emph{irrelevant} parts, with coupled equations of motion. 
By formally solving the dynamics of the second part, the evolution of the system can be written as a time-non-local differential equation 
of the form
\begin{equation}
\dot \rho(t)=-i {\cal L}\rho(t) +  \int_0^t d\tau
{\cal K}(t-\tau)\rho(\tau),
\label{eq:gqme}
\end{equation}
involving the convolution of the system density operator with the memory kernel ${\cal K}$.
The complexity of the problem is thus transferred to the computation of the memory kernel, which 
encodes how much the
interaction with the environment causes the dynamics of the system to depend on its past states. The memory kernel
can be determined in terms of time-dependent bath correlation functions~\cite{Shi2003kernel}, or from the reduced density operator for the system for a complete set of initial conditions~\cite{Montoya2016}.
In many practical scenarios, this memory kernel decays quickly, such that it becomes possible to compute it from approximate or quasi-exact simulations of the complete dynamics for finite times.
In such cases, the GQME approach provides an efficient approach to simulate the dynamics of the reduced system.

Since the process tensor maintains all relevant information about the temporal correlations in the bath, it is logical to expect that it allows retrieving the memory kernel information.
One may also expect that a finite memory of the GQME kernel should correspond to a less correlated influence functional, which, in turn, should also be easier to represent as a MPO. 
A connection between both approaches was first hinted at in Ref.~\cite{Golosov}, shown explicitly in 
Ref.~\cite{Cerrillo2014NonMarkovian}, and later generalized in Ref.~\cite{Pollock2018tomographically}, by explicitly constructing the discretized memory kernel in terms of \emph{transfer tensors}  that can be extracted from a knowledge of the dynamical maps---as provided by the influence functional---at different times.
The transfer tensors $T_{n,k}$ ($k=0,\ldots n-1$) are superoperators acting on the state of the system, which provide a decomposition of the dynamical map for time $t_n$ ${\mathcal E}_n$, such that the
state of the system at a given time is expressed as a sum of propagations from past states, as
\begin{equation}
    \rho(t_n)={\cal E}_n \rho(0)=\sum_{k=0}^{n-1}T_{n,k} \rho(t_k).
\end{equation}
This can be interpreted as a discrete integral form of the Nakajima--Zwanzig equation, where the memory kernel can be read out from the terms that encode the non-local in time propagators ($k<n-1$)~\cite{Cerrillo2014NonMarkovian}. 

A similar connection can be obtained from the Quasi-Adiabatic propagator Path Integral (QuAPI) framework~\cite{Makri1992,Makri1995Tensor1,Makri1995Tensor2}, which provides a numerically exact way to evaluate the discretized path integral for a system coupled to a harmonic bath. The corresponding influence functional can be expressed as a product of terms connecting two different times.
In the Small Matrix Path Integral (SMatPI) approach~\cite{Makri2020Small,Makri2021Small}, each of these terms is split into a factorized component and a smaller remainder that contains the non-local in time correlations.
This allows the propagator of the system to be rewritten as a sum with the same structure, 
\begin{equation}
    {\cal E}_{n}=\sum_{m=1}^{n-1}M^{nm} {\cal E}_m+M^{n0},
\end{equation}
where the matrices $M^{nm}$ encode the genuine time correlations at distance $n-m$. Truncating the sum to a maximal length $r_{\max}$ thus corresponds to a finite memory of the GQME kernel (see also Ref.~\cite{Chatterjee2019methods}).
The complete equivalence between memory kernel and influence functional was more recently demonstrated for linear interactions of the system with a Gaussian bath in Ref.~\cite{Ivander2023unified}, which explicitly connected the components of the memory kernel to the terms of the functional, and furthermore showed how to extract the spectral density of the bath from the influence functional, provided the free Hamiltonian of the system is known~\footnote{Note that there exists some controversy over the derivation of this connection with respect to the nature of time discretization.  See~\cite{Makri2025,LindoyLee2025}}.

\section{Comparing and Benchmarking available approaches}
\label{sec:benchmarking}

\begin{table*}[ht!]
\caption{Summary of representative publicly available software packages for non-Markovian open quantum systems.}
\label{tab:packages}
\begin{tabular}{ ||p{4.2cm}|p{3.5cm}|p{9.25cm}||  }
 \hline
 Package& Algorithm &Comments\\
 \hline
   OQuPy~\cite{FuxOQuPy}   & \multirow{4}{3.5cm}{PT-MPO}    & -Treats Gaussian bosonic baths. Can use calculated PT-MPO for mean-field limit of one-to-all coupled systems, and for chains of coupled systems.\\
   
   ACE~\cite{ACEgauger}  &    & -Treats general non-Gaussian (uncoupled) bath modes with ``automated compression of environments'' (ACE) algorithm.\\ \hline
   
 PathSum~\cite{PathSum} & QuAPI and related algorithms  & -Employs iterative form of QuAPI and modular path integrals (MPI) and decompositions such as SMatPI to reduce memory overhead.\\
 
 QuantumDynamics.jl~\cite{QDBose} & General IF approaches & -Approximate and exact frameworks, tensor-network IF as well as QuAPI formalisms.\\ \hline
 
 MPSDynamics~\cite{MPSDynamics}    &T-TEDOPA& Thermalized dynamic MPS-based TEDOPA algorithm.\\ \hline
 
 HEOM-QUICK~\cite{HEOM-QUICK} & \multirow{3}{3.5cm}{HEOM}  & -Treats fermionic impurity problems/baths.  \\
 HierarchicalEOM.jl~\cite{HierarchicalEOM.jl} &  & -Treats simultaneous or separate bosonic or fermionic baths.  \\ 
 MPSQD~\cite{Shi2024} & & -MPS-based HEOM code with direct MPS/MPO functionality. \\ \hline
Heidelberg MCTDH~\cite{HeidelbergP} & \multirow{4}{3.5cm}{ML-MCTDH} & \multirow{4}{7cm}{-Codes with ML-MCTDH and in some cases other functionalities.}\\
Quantics~\cite{Quantics}& &\\ 
RENORMALIZER~\cite{Renormalizer}& & \\ 
pyTTN~\cite{pyTTN} & & \\ \hline
MesoHops~\cite{MesoHops} & HOPS & -Adaptive HOPS code. \\
 \hline
\end{tabular}
\end{table*}
Having outlined a variety of interconnected methodologies centered around the concept of the process tensor for the description of the non-Markovian dynamics of open quantum systems, a natural question arises as to which approaches are best tailored to specific problems of current interest.  Of course this question is rather open-ended: it depends on goals (e.g. numerical efficiency versus flexibility and generality) as well as the specifics of the physical problem under consideration.  We do not draw strong conclusions here, and we merely point out (in an illustrative and non-exhaustive manner) the myriad of useful software packages that exist, outline ways these and future codes can be put to the test, and conclude with a discussion of important forefront physical problems whose dynamical behavior is difficult to capture and which can potentially benefit from the use of the methods outlined herein.

We begin by highlighting in Table~\ref{tab:packages} a few packages which exist that enable a numerically convergent treatment of non-Markovian dynamics in a variety of different open quantum system contexts.  The list we provide is far from exhaustive, and in many cases the packages mentioned below employ more algorithmic techniques than the highlighted formalism.  Influence functional methods like QuAPI, tensor network approaches like TEDOPA, and HEOM can be connected to the more general concept of the process tensor, as discussed earlier in this Perspective.  In addition, we include the multi-layer multi-configurational time-dependent Hartree (ML-MCTDH) approach.  ML-MCTDH employs a tree-tensor network description of the wave function directly, for which the connection to process tensor or influence functional approaches has not yet been put forward.  Unlike the algorithms employed in the codes mentioned above, ML-MCTDH uses a discretized bath with mode frequencies and couplings fit from a continuous function in cases where the model is described by a smooth spectral density. In such cases additional convergences tests must be performed with respect to the mode spacing but may prove to be an advantage in cases where the reservoir contains discrete modes.

A natural question which arises concerns the efficiencies of the algorithms and codes mentioned above, as well as those not discussed here.  First, we emphasize that there is a distinction between the potential performance of a theoretically formulated algorithm and the efficiency of the code that employs it, which depends on details of optimization and other considerations.  While obvious, this point complicates the relative comparisons of advantage of given methodologies. Accepting this reality, one may still attempt to understand under what physical circumstances it is preferable to use a given approach over another one.  Even here this will depend very much on the problem and the observables. A major conceptual benefit of the process tensor approach is that it unifies many of the methodologies discussed herein.  This unification exposes more general routes to consider how to optimize tensor network architecture and tensor contractions for a given problem.  

To illustrate the subtlety of the question of the dependence of efficiency on the target observables, consider the case of the calculation of multi-time observables in open quantum systems.
Once the process tensor has been constructed, it will still take approximately $N^{m}$ independent contractions to extract $m$-time correlators over N time steps---this scaling is the number of output values to be calculated, as $m$-time correlations lead to $m$ independent time indices to vary.
The number of independent propagations for other methods that do not explicitly construct the process tensor will be the same, but the cost of each propagation would be higher.
Unlike with other approaches, process tensors have an upfront construction cost.
If one uses the original~\cite{Jorgensen2019Exploiting} approach to calculating the process tensor, as opposed to the time-translational invariant process tensor (TTI-PT) approach~\cite{Link2024Open}, then the upfront costs scales as $O(N K)$ for memory truncation after $K$ steps, or $O(N^2)$ with no memory truncation.
This statement assumes the typical case where the bond dimension of the MPO saturates at large $N$, otherwise the scaling with $N$ would be larger.
In this case, whether the up-front cost exceeds the cost of calculating desired outputs will then boil down to whether $m>2$ (without memory truncation) or $m>1$ (with truncation).
When the output contraction cost exceeds the upfront cost, one would expect the the PT method to always be more efficient due to the reduced cost of calculating the outputs.
In other cases which method is most efficient will depend  on the runtimes of different methods for a single propagation step.
Moreover, one may then in principle use the  eigenspectrum of the TTI-PI to essentially calculate any correlation in constant time.
These considerations will play a role in determining how to tailor a given approach to a given set of observables.

Simple impurity problems like the spin-boson model with specific spectral densities have been “go-to” benchmark problems for over thirty years since the advent of numerically controllable deterministic QuAPI calculations~\cite{Makri1995Tensor1,Makri1995Tensor2} and the development of real-time quantum Monte Carlo approaches~\cite{ChandlerMak, EggerMak}.  Interestingly, while hundreds of such benchmark papers have been published showcasing the ability of a given approach to converge to a “numerically exact” result for set of parameters of a given spin-boson problem, we know of no attempt to broadly compare many different methods on the same problem. In general the simple spin-boson problem can be characterized by the spectral density $J(\omega)=A\omega^{s}\Theta(\omega, \omega_{c})$ where $A$ characterizes the strength of the coupling of the two-level system to the environment and $\Theta(\omega, \omega_{c})$ is a function that specifies the spectral width of the environment with a high-frequency cut-off $\omega_{c}$~\cite{Leggett1987dynamics}.  Some prominent examples include the Ohmic case $s=1$ which is appropriate to model "bosonic" electron-hole pairs in a metal~\cite{Leggett1987dynamics}, the super-Ohmic case $s=3$, appropriate to model the coupling to acoustic phonons with the deformation potential approximation~\cite{SilbeyHarris}, and the sub-Ohmic case $s<1$, appropriate for some quantum information applications~\cite{LeHur}.  In addition, more realistic problems often contain sharp spectral features on top of a smooth background spectral density which can model, for example, discrete intramolecular modes that couple to the system degrees of freedom.  Such features can present a challenge for methods that employ memory truncation.

A reasonable starting point for comparison benchmarks would be to consider the most challenging of the simple spin-boson cases, namely the sub-ohmic model.  This model exhibits a line of zero temperature localization critical points as well as prominent non-Markovian dynamics of the system’s spin degree of freedom at low temperatures~\cite{Bulla}, it would appear to be a reasonable first test case even though the model does not contain many physical ingredients of interest.  A convincing benchmark study would consider the performance of the methods described above with respect to efficiency across the parameter space “phase diagram” associated with axes of all independent energy scales (e.g. temperature, system-bath coupling, bath cutoff frequency).  Various observables (single-point versus multi-point correlations) can be considered.  Despite the fact that a benchmark of this nature is rather circumscribed, it would require a sizable effort.  We feel that this effort is valuable and will be of significant benefit for the future development and design of process tensor methodologies as well as methods for the exact simulation of non-Markovian processes in general.  A further step in the program of benchmark studies should include spectral densities with sharp spectral features which represent individual modes with long lifetimes.  Consideration of these cases may lead to useful hybrid methods that combine some of the advantages of distinct approaches,  such as, e.g., process tensor methods with the ML-MCTDH approach.

\section{Using Process Tensors}
\label{sec:Future:Apps}

Having discussed various ways process tensors can be calculated, in this section we turn to how process tensors, once compressed into tensor networks, can enable certain calculations that would otherwise be prohibitively costly or difficult to formulate. 
In this section we outline some types of calculations the process tensor formalism enables.
The next section discusses examples of physical problems which illustrate where process tensors may be particularly powerful approaches.

As noted above, process tensors are particularly valuable when multiple calculations need to be performed on the same open quantum system.
In cases where all that is needed is to perform time evolution and calculate single-time expectation values $\langle O(t)\rangle = \text{Tr}[\rho(t) \hat O]$, a process tensor calculation 
may not be the most efficient way to perform such a calculation,  
although a process tensor may still be more efficient if approximating it numerically is easier than simulating the full time evolution of the observable, as would be the case when the temporal entanglement grows slower than the spatial one~\cite{cerezoro2025}. 
However, there are many scenarios where one can frame a calculation as first producing, and then repeatedly using, a process tensor.  Examples of this include:
\begin{description}
    \item[Time-dependent fields]
    The simplest extension to the single time evolution process mentioned above is cases in which one wants to run a simulation many times, with either varying static parameters, or with some arbitrary time-dependent Hamiltonian applied to the system, such as describing the result of a particular pulse sequence applied to the system by a time dependent field.     This idea also forms the foundation for two further applications:

    \item[Quantum control and optimization] The idea of optimal quantum control~\cite{glaser2015training,koch2022quantum} is to design control fields that produce a desired output state in a realistic (and thus open) quantum system.   In addition to process tensors allowing one to calculate the result of many different pulse sequences, they can also allow one to calculate the gradient of the final state with respect to the time dependence of the pulse~\cite{Butler2024}.  This allows one to use methods such as the Gradient Ascent Pulse Engineering (GRAPE)~\cite{khaneja2005optimal} algorithm to produce the optimal pulse sequence.

    \item[Self-consistent fields] In addition to external imposition of time-dependent fields, one can also consider combining the process tensor with mean-field approaches, in which each part of the system sees a self-consistent field resulting from other parts of the system~\cite{Chaikin1995PrinciplesPhysics,fazio2025manybodyopenquantumsystems}.
    This can be seen as a special case of time-dependent fields, as discussed above, but where the time-dependence is not externally imposed.
    
    The combination of process tensor and mean-field methods provides a potential route to approximately describe many-body non-Markovian problems.
    This could involve a lattice of sites with a non-Markovian dissipation on each lattice site, and mean-field decoupling between each site~\cite{Fisher1989BosonTransition,krauth1992gutzwiller}.
    In such a case, mean-field theory is an approximation, and only valid when the lattice is effectively high dimensional.
    Another example is models with one-to-many coupling, such as the Dicke model describing many emitters coupled to a central photon mode~\cite{Hepp73equilibrium,DickeReview}.
    Here mean-field theory is expected to become exact as the number of emitters becomes large~\cite{Carollo2021Exactness}.
    We discuss this further in Sec.~\ref{sec:Future:Apps:DIcke} below.

    \item[Multi-time correlations] This refers to cases where one wants to calculate a multi-time expectation $\langle O_1(t_1) O_2(t_2) \ldots O_n(t_n) \rangle$ (we restrict here to time-ordered correlations where $t_{n-1}<t_n$).
    These can be directly calculated from a process tensor by including the appropriate operators $\hat O_n$ at time step $t_n$ along with the pure system evolution operator. 
    For two-time correlations this can be used to extract response functions used for linear spectra~\cite{FowlerWright2022} or to probe the fluctuation-dissipation theorem~\cite{Fux2023Tensor}.
    With four-time correlations this can be used to model~\cite{deWit2025process} two-dimensional coherent spectroscopy~\cite{Mukamel,cho:2008,collini:2021}.

    \item[Process tensor tomography]  A rather different use of process tensors occurs when considering what measurements of a non-Markovian system need to be made in order to understand its general evolution.  Such an approach builds on the idea of ``quantum process tomography''~\cite{Poyatos1997,chuang1997prescription} where one seeks to characterize a quantum process by a sequence of state preparation and measurements.  This has been extended to characterizing a non-Markovian open quantum system by tomography to measure the process tensor~\cite{Luchnikov2020Machine,Guo2020Tensor,White2020Demonstration,White2022}.
    
\end{description}

For a fuller summary of different kinds of calculations that can be done efficiently with process tensors, see Ref.~\cite{FuxOQuPy}.

\section{Future directions}
\label{sec:future}

We hope the discussion above has made clear the potential power of the process tensor approach, and tensor-network representations of the process tensor, as a practical tool for modeling non-Markovian quantum dynamics.
To conclude this perspective, we discuss some directions that we expect are likely to lead to future developments in this field.
We divide this into two, Sec.~\ref{sec:future:apps} focusing on applications where process tensor methods may be usefully applied, and Sec.~\ref{sec:future:methods} focusing on methodological improvements that may enable a broader range of systems to be explored.   
We note that some methods may be more or less suitable for some physical problems, and so there is not always a sharp divide between these two sections.
The applications of process tensors  discussed below are not intended to be comprehensive. Instead, they are selected to illustrate the range of ways process tensors can be used. 

\subsection{Applications}
\label{sec:future:apps}

\subsubsection{Collective Behavior in Polaritonic Systems and Quantum Optics Settings}
\label{sec:Future:Apps:DIcke}
There has been a surge of recent interest, both experimentally and theoretically, in materials and atomic and molecular assemblies placed inside an optical cavity, where strong coupling between the confined electromagnetic field and the confined matter leads to the formation of hybrid light-matter (polaritonic) states and the potential induction of emergent collective phases~\cite{carusotto2013quantum,sanvitto2016road,Keeling202BEC,Basov2021polariton}.  

One example where process tensor methods have already been used is the investigation of the potential role of dissipation and dephasing in the Dicke model~\cite{DickeReview}.  This model describes a system of N two-level systems coupled to a common cavity mode, and as a closed model may exhibit zero temperature (quantum) and finite temperature (classical) phase transitions to a ``superradiant state''~\cite{Hepp73equilibrium}.  It is now understood that if one creates an ``open'' Dicke model, where a Markovian (Lindblad) dephasing bath is connected to each two-level system, then the transition to a superradiant phase is destroyed~\cite{Demler,KirtonKeelingDecay}.  This transition may be restored if, in addition to Markovian dephasing, Markovian population relaxation is also included.  What happens when each two-level system couples to a general non-Markovian bath is unclear, except in some limiting cases~\cite{Nagy}.
Using process tensors, a model of many molecules collectively coupled to a single photon mode and strongly coupled to many individual vibrational modes has been 
This system,  which serves as model for organic polaritons, has been explored both using mean-field theory~\cite{FowlerWright2022}, as well as with extensions beyond mean-field theory~\cite{mu2025beyond}.

A more challenging but related problem is the modeling of putative collective alteration of chemical reaction rates of assemblies of molecules within solution and confined inside a cavity, an effect which has been suggested by recent high profile experiments, as reviewed in Ref.~\cite{Ribeiro2018Polariton,Herrera2020Molecular, GarciaVidal2021,Simpkins2021Mode,Mandal2023Theoretical,Nelson2024More}.  This problem is challenging because even under the mean-field approximation where an N-molecule system is transformed to a one-body problem, the single “body” may consist of many vibronic energy levels coupled simultaneously to a solvent bath as well as to a (potentially lossy) cavity.  A recent study has carried out such simulations using the HEOM approach, finding that while rates are not altered, shorter time non-exponential dynamics may be markedly altered by non-equilibrium preparation effects~\cite{Lindoy2024}.  However, this study made a number of simplifying approximations, including the neglect of direct energy transfer pathways between molecules as well as the invocation of the single mode approximation for the cavity.  Relaxing these approximations presents challenges for numerical approaches.  We believe that a promising future direction for this problem would be the use of process tensor methods.

\subsubsection{Energy Transfer Dynamics in Complex Biological Systems}
\label{sec:Future:Apps:Bio}
Another longstanding problem of interest where the process tensor methods outlined in this perspective can be of great utility is the description of the non-Markovian dynamics of energy transfer in biological systems such as photosynthetic pigments~\cite{scholes2011lessons,huelga2013vibrations,lambert2013quantum,wang2019quantum,Cao2020Quantum}.  The dynamics of these systems has been elucidated by a variety of sophisticated spectroscopic techniques, which has enhanced our mechanistic understanding of this process in a range of different photosynthetic complexes.  A long-standing question in this field is the role played by coherences, whether electronic or vibrational, in the energy transduction process.  Although simplified models have been employed to describe non-Markovian quantum dynamics in photosynthetic systems using a variety of approaches (TEDOPA~\cite{Caycedo-Soler}, the modular path integral (MPI) approach--a technique conceptually related to QuAPI which enables the efficient treatment of extended systems with local couplings~\cite{Kundu2021Efficient,Kundu2022Tight}, and HEOM~\cite{Ishizaki}, to name a few), realistic models of photosynthetic complexes contain many states coupled to vibrational reservoirs that contain spectral features over a broad range of frequencies in addition to sharp intramolecular modes~\cite{Coker}.  Furthermore, couplings may not be linear, especially in cases of intramolecular vibrational modes, which themselves may display anharmonicity, thus requiring a more realistic description than that found in the simple multi-state spin-boson Hamiltonians that have been frequently employed.  Lastly, many experiments use powerful multi-dimensional spectroscopies to probe these systems.  The theoretical calculation of such signals thus requires the ability to treat multi-point correlation and response functions.  We believe process tensor approaches can be of great utility for the study of these problems.

\subsubsection{Coherent multidimensional spectroscopy of biological systems}
\label{sec:Future:Apps:Spectroscopy}
The first direct evidence of the role of coherence in photosynthetic energy transfer came from two-dimensional spectroscopy, and two-dimensional electronic spectroscopy in particular has provided a wealth of information related to energy transfer in these systems~\cite{engel2007evidence}.  The output of a multidimensional spectroscopy experiment is a Fourier map of a multi-time correlation function of a system which arises from specific interventions of controllably-timed external pulses~\cite{Mukamel,cho:2008,collini:2021}. Although very powerful, the interpretation of multidimensional spectral features in complex systems is subtle and difficult.  For example, long-lived oscillatory features in two-dimensional spectra may arise from quantum coherence between electronic states, semiclassical vibrational motion on one or more vibronic surfaces, or some combination of both~\cite{Roel}.  These features are more prominent at low temperatures, where coherence of all types is longer lived, and non-Markovian relaxation is more pronounced; however, they may be pronounced even near room temperature due to sharp spectral features associated with particular environmental modes that strongly couple to the electronic states of the system.

In general, modeling the multi-time correlation functions associated with experimentally-determined multidimensional spectra is difficult, but process tensor methods are ideally suited for calculating such quantities in a very general way~\cite{deWit2025process}.  As noted earlier, multi-time correlation functions can be naturally calculated in a time-translationally invariant process tensor MPO (TTI-PT-MPO) framework~\cite{Link2024Open} by contracting a sequence of operations corresponding either to the insertion of temporally separated probe operators or field-free propagation, depending on the time interval. The time evolution between pulses can be calculated via divide-and-conquer methods or eigenvector projection approaches, which can greatly reduce the scaling cost, essential for the modeling of multidimensional spectroscopy associated with energy transduction in biological systems such as photosynthetic complexes.  Optimal protocols will depend on the complexity of the underlying model system and the particular pulse sequence, among other factors.

An untapped advantage of the use of process tensor methods is that distinct pulse sequences can be used to calculate multidimensional spectra without the need to recompute the process tensor every time~\cite{Fux2021Efficient}.  This feature means that optimal pulse sequences can be efficiently designed {\em in silico} to target specific features in the multidimensional spectrum and gain insight into the connection between these features and the underlying microscopic components of the Hamiltonian and the particular dynamical processes from which they emerge~\cite{deWit2025process}.  

Another potential direction is to use process tensors as a tool to explore the potential offered by entangled photon spectroscopy~\cite{schlawin2017entangled,schlawin2018entangled,eshun2022entangled}.  In the process tensor framework, as well as processes involving either unitary evolution or external classical pulses, one can directly introduce quantum states which interact with the system at controlled times.  As such, the process tensor can provide a natural framework to perform calculations of spectroscopy resulting from interactions between the system, the environment, and an entangled photon source.

The further development and application of tools based on process tensor methods will be invaluable for the interpretation of two-dimensional and related spectra of systems such as photosynthetic complexes.

\subsubsection{Mobile impurity problems}
An area of current research where there is as yet little connection to process tensors is the area of mobile impurities, and the formation of Bose- and Fermi-polarons~\cite{Massignan2014,Levinsen2015Stronly,Schmidt2018universal,grusdt2024impurities,parish2025fermi}.
Such problems arise in various contexts including ultracold atomic gases,  exciton-polariton condensates, and electrically doped semiconductors.
They describe cases in which an impurity (another atom, or some form of optical excitation) interacts with a medium that can deform in some way, allowing the mobile excitations to be dressed by a cloud of excitations of the medium, analogous to the dressing of a charge by deformations of the lattice in a standard polaron.

Most work on such systems has not started from the open quantum system framework, but has instead sought to write down wavefunctions in the full space of the impurity and the medium.
However, in any case where the relevant observables can be cast in terms of properties of the impurity, an open-systems perspective ought to be possible.
The primary challenge in such cases is that if the impurity is mobile, the state space it can explore is particularly large.

Very recently, there has been work~\cite{Gao2025summing} showing how process tensors can be derived for mobile impurities in the Holstein model.
This work made use of the flow equation approach~\cite{Wegner1994flow} to derive a differential equation determining how the process tensor evolves with system-bath coupling strength.
This equation can then be solved by standard tensor network methods, such as the Time-Dependent Variational Principle (TDVP)~\cite{Haegeman2011}.
This work both suggests the possibility of methods that can be extended to other mobile impurity problems, as well as suggesting another distinct approach to deriving process tensors.

\subsection{Methods}
\label{sec:future:methods}

\subsubsection{Beyond the MPO ansatz}

The discussion in previous sections has focused on the use of the MPO ansatz for the description of the process tensor.
This approximate representation has the practical advantage of the availability of very efficient numerical methods, and it allows an almost direct physical interpretation of the bond dimension in terms of the environment memory.
At the same time, the two-time correlations in a MPO process tensor necessarily decay exponentially with the time difference. 
This ansatz can efficiently capture the relevant phenomena of multiple physical scenarios, as numerical studies have demonstrated, but it does not exhaust the possible structure of environment correlations.
Alternative representations may be useful for the construction of process tensors with higher degrees of temporal correlation associated with particular geometries of the environment.

This possibility was studied in Ref.~\cite{Dowling2023process} which introduced the idea of the \emph{process tree}, a process tensor with the geometry of a tree tensor network (TTN)~\cite{Shi2006ttn}. The process tree is parametrized by maps that introduce finer timescales as moving further from the root.
It inherits properties from the TTN, such as the possibility to efficiently compute correlations, due to its causal structure, and the support of critical (power-law) correlations (in time).
The results in Ref.~\cite{Dowling2023process} show that this ansatz can efficiently capture the correlations in the critical spin-boson model.  A limitation is that the results of Ref.~\cite{Dowling2023process} were found by fitting a TTN to the already-known process tensor of a particular model.  An open question is whether one can directly proceed to find the TTN construction from a microscopic model of the system-bath coupling.  
Exploring the details of the TTN structure of MCTDH~\cite{lindoy2021time,lindoy2021time2, larsson} may be a potential avenue for exploration here.

Finally, we mention the possibility that process tensors could also be represented through other compressed representations, such as restricted Boltzmann machines or other forms of neural networks.  
A restricted Boltzmann machine is a shallow neural network, with one visible layer and one hidden layer of neurons.  
It has been used as a variational ansatz for the quantum wave function of the ground state of a many-body problem~\cite{Carleo2017SolvingNetworks}, as well as for a description of open quantum systems
\cite{Torlai2018Latent,Nagy2019variational,Hartmann2019neural,vicentini2019variational}.
In both cases, it serves as a representation of a high-order tensor, which can be variationally optimized to minimize a Hamiltonian or find the stationary state of a master equation.
The same approach could potentially be used as an alternate representation of the process tensor.

\subsubsection{General Environments}  One of the major advantages of process tensor methods is their generality.  
There always exists a process tensor to represent the effect of any environment on any system.
As such, in principle, they provide a path to go beyond the simple models of linearly-coupled Gaussian baths, and can formally treat situations involving non-linear coupling, correlated initial conditions, and non-Gaussian (anharmonic) baths composed simultaneously of fermionic, bosonic and spin degrees of freedom.  The ability to describe non-Markovian dynamics in such systems is a frontier problem for which some strategies are now emerging.  For example, the ``automated compression of environments'' (ACE) approach~\cite{ACEgauger} can produce a PT-MPO for any environment that is made up of independent (not mutually interacting) degrees of freedom.

Another interesting direction to pursue are approaches to extract the process tensor from explicit representations of environments beyond one dimension.
In Sec~\ref{sec:PT:unifying:chain} we discussed how the process tensor could be directly extracted from one-dimensional chain representations of environments.
One may also consider whether this approach can be extended to higher dimensional geometries.
This is relevant if one seeks to use process tensors to describe the dynamics of impurities in a many-body interacting system, such as if the problem at hand is the study of local observables in a two-dimensional lattice model.
In this scenario it is possible to construct a spatio-temporal TN analogous to the one described in Sec.~\ref{sec:PT:intro}.
Reference~\cite{Park2025twodimensional} showed that if the environment lattice has the geometry of a tree, i.e. there are no loops, the influence functional can be approximated by a product of MPS, which can be found efficiently via an iterative belief propagation procedure.
For lattices with loops,  Ref.~\cite{Park2025twodimensional} suggested an approximation scheme based on a cluster expansion of the influence functional. 

While the methods mentioned here and above mean that viable approaches exist for many different forms of environment, further extensions to other forms (e.g. interacting and anharmonic modes) remains a challenge.
Addressing this will greatly enhance the scope of process tensor approaches to non-Markovian quantum dynamics in general settings.

\subsubsection{Large and many-body systems}
Another important challenge for future research is to model non-Markovian dynamics for systems with large Hilbert spaces.  
Increasing the size of system Hilbert space is a challenge for both Markovian and non-Markovian problems alike, but the larger overhead for non-Markovian methods makes the combination all the more difficult.
Indeed, as a result of this, most work on non-Markovian open systems has been restricted to systems represented by a small number of levels or simple cases such as non-interacting bosons.

Some approaches have already been introduced which allow a larger Hilbert space to be considered.  
One significant simplification arises when the system operator which couples to the environment has only a limited number of non-zero eigenvalues.  
Using a method first discussed in the context of QuAPI, Ref.~\cite{Cygorek2017nonlinear} noted that for cases where a Hermitian system operator $\hat X$ couples linearly to a harmonic-oscillator bath, the number of required terms describing the system-bath coupling depends on the number of distinct values of eigenvalue differences $x_\lambda-x_\lambda^\prime$, where $x_\lambda$ is an eigenvalue of $\hat X$.
This reduction can be particularly useful when considering cases where only a subpart of the system couples to the environment---such as a few-level atom in a cavity where the environment couples only to the atom, or where a $\hat X$  is a symmetric combination of system operators such as in superradiance~\cite{cygorek2024sublinear}.

More recently, an alternate approach has been proposed that is highly suited for cases where the system operator $\hat X$ has many eigenvalues, providing an almost-smooth continuum of values.  In that case, one may write the influence function in terms of interpolating functions, reducing from needing to index the Hilbert space to instead indexing the number of interpolating functions used. This approach, Quantum Accelerated Stochastic Propagator Evaluation (Q-ASPEN), is discussed in Ref.~\cite{Grimm2024stt}.
For particularly large Hilbert spaces, there can be an advantage in using stochastic unravelling techniques~\cite{Plenio1998quantumjump}, where one evolves an ensemble of wavefunctions of the system, rather than the density matrix.  Non-Markovian versions of this have recently been explored by Ref.~\cite{mueller2025nuhops}, however at present, there is no direct way to relate such wavefunction-based approaches to process tensor paradigm.

One particular physical context where such questions become relevant is that of multi-orbital and multi-impurity Anderson models~\cite{Bulla2008rmp,Gull2013Numerically,Cohen2015Taming}.
Recent work has pointed the way towards the formulation of tensor-network based influence functional methods for the Anderson model~\cite{Thoeniss2023Efficient,Ng2023Realtime,Sonner2025,nayak2025dmft}.  Additional recent work has suggested how such approaches may be optimized.  HEOM formulations to describe long-time dynamics in this model have existed for some time, and implementations of HEOM for such fermionic bath models can be found in several of the packages mentioned above~\cite{HEOM-QUICK}.  The generality of the process tensor concept, and the fact that it can connected the tensor network influence functional approach with the HEOM approach, suggests that this framework can be of great utility for paving the way to formulate optimized ways of dealing with Anderson type models.  While the single impurity Anderson model is already tractable with currently available methodologies, the more challenging case of models with multiple orbitals per site and/or multiple impurity sites is largely unexplored and it remains unclear to what degree the bond dimensions for tensor network approaches can be kept manageable during the course of time evolution.  These more complex models are of great importance in condensed matter physics and materials science, and often arise in cluster DMFT studies of complex solids~\cite{Georges1996Dynamical,Aoki2014Nonequilibrium}.  It should be noted that even sophisticated imaginary time quantum Monte Carlo impurity solvers face a difficult sign problem when multiple orbitals and multiple impurity sites are involved~\cite{Gull2011Continuous}.

Another broad context where is necessary to model large systems is where the ``system'' itself consists of many coupled subsystems.  
These naturally arise when considering quantum many-body problems such as lattice models of interacting spins, or variants of the Hubbard and Bose-Hubbard models~\cite{Bloch2008Many-bodyGases}.  
As noted in the introduction, tensor network methods have already been extensively applied to approximating the structure of spatial correlations and entanglement in such models
when isolated~\cite{Schollwoeck2011DensityMatrix,Orus2014Practical} or subject to Markovian environments~\cite{Weimer2021simulation}.
This suggests that it may be possible to describe such systems by combining tensor network approaches in both space and time (memory); preliminary work toward this goal has been discussed in Ref.~\cite{Bose2022multisite,Fux2023Tensor}.

There are several contexts in which it may be of interest to describe such many-body non-Markovian open quantum systems.
One context relates to recent interest in driven-dissipative many-body systems, and the phase transitions and collective behavior seen in such systems~\cite{sieberer2023universality,fazio2025manybodyopenquantumsystems}.
To date, most discussion of such behavior has assumed Markovian dissipation.
One may however anticipate that different critical behavior may arise from structured environments, as has already been noted in some specific examples~\cite{Nagy}.

The other broad context where such problems may arise is in the context of quantum materials~\cite{keimer2017physics}, seeking to include e.g. vibrational modes or external fermionic reservoirs as realistic non-Markovian environments.
In particular, such questions become relevant when considering driven quantum materials, or quantum materials strongly coupled to optical cavities~\cite{Schlawin2022Cavity,Basov2025polaritonic}.
The mobile quantum impurity problem mentioned above is one such problem in this context where methods tackling large Hilbert spaces become relevant.
In addition, in the context of quantum materials, there is an obvious connection back to quantum impurity problems and DMFT as discussed earlier.
These introduce the additional feature of self-consistency in cases where the environment dynamics should be given by the process tensor found for a given impurity. As noted in Sec.~\ref{sec:Future:Apps}, in principle such self-consistent approaches are compatible with PT approaches.
As such, there is considerable scope to use PT approaches as impurity solvers~\cite{Thoeniss2023Efficient,Ng2023Realtime,Sonner2025,nayak2025dmft} for DMFT.
There may also be ways to relate the PT-MPO to other approaches that have been recently developed for such problems~\cite{NunezFernandez2022Learning}, based on using matrix product states to efficiently evaluate the multidimensional integrals arising from diagrammatic expansions.

As commented at the start of this perspective, any problem in which one is interested in measuring observables on only one subset of degrees of freedom can be framed as an open quantum system problem.
Ultimately, increasing the technical capability of process tensor methods---in terms of system size, form of environment, and efficiency of representation---may allow such methods to be directly applied to studying many-body physics in quantum materials, significantly expanding the scope of the open quantum system paradigm.

\begin{acknowledgments}
The authors would like to thank
Joel Beckles,
Garnet Chan,
Alex Chin,
Moritz Cygorek,
Roosmarijn De Wit,
Paul Eastham,
Piper Fowler-Wright
Gerald Fux,
Erik Gauger,
Peter Kirton,
Brendon Lovett,
Andrew Millis, 
Kavan Modi,
Nathan Ng, 
Gunhee Park, 
Olivier Parcollet,
and
Aidan Strathearn
for collaborations and discussions on topics related to this work.
This perspective arose from a meeting hosted at the Simons Foundation's Flatiron Institute Center for Computational Quantum Physics (CCQ); we are grateful for sponsorship for that meeting from CCQ and the EPSRC International Quantum Tensor Network (EP/W026953/1).
JK acknowledges support for related work from EPSRC (EP/T014032/1).
He would also like to thank Gerald Fux for providing a version of figure 1.
MCB acknowledges partial support from the DFG (German Research Foundation) under Germany's Excellence Strategy -- EXC-2111 -- 390814868, Research Unit FOR 5522 (grant nr. 499180199) and TRR 360 - 492547816, 
and by the EU-QUANTERA project TNiSQ (BA 6059/1-1).
DRR acknowledges support from the U.S. Department of Energy, Office of Science, Office of Advanced Scientific Computing Research, Scientific Discovery through Advanced Computing (SciDAC) program, under Award No. DESC0022088.  
He would also like to thank Nathan Ng for assistance on this manuscript.
\end{acknowledgments}

\appendix

\section{Appendix: Details of Free Boson Process Tensors}
\label{app:details}

Section~\ref{sec:PT:computing} reviewed an exact expression for the process tensor of a system of 
free or Gaussian bosons. Here we give more details of some of the quantities appearing in that section.

The process tensor can be written as a product of exponentials with factors $\eta_{i-j}$ appearing in the
exponents. These $\eta_{i-j}$ factors are given in terms of the environment correlation function $C(t)$ as
\begin{align}\label{eq:defn_eta}
\eta_{i-j} = \begin{cases}
                 \int_{t_{i-1}}^{t_i} dt' \int_{t_{j-1}}^{t_j} dt'' \ C(t' - t'') &  (i\neq j) \\
                  \int_{t_{i-1}}^{t_i} dt' \int_{t_{j-1}}^{t'} dt'' \ C(t' - t'') & (i = j) 
              \end{cases}
\end{align}
where we have assumed a time translation invariant environment correlation function $C(t_1,t_2) = C(t_1-t_2)$.
Note that $\eta_{i-j}$ is indeed a function only of the difference of index $i-j$, even though the definition in Eq.~\eqref{eq:defn_eta} appears to depend on $i,j$ separately---one may readily see that replacing $i,j \to i+k,j+k$ leaves the expression unchanged as long as the environment correlation function $C(t_1-t_2)$ is time-translation invariant.
The $\eta_{i-j}$ therefore capture correlations of the environment averaged over intervals of time between
the discretized time points corresponding to process tensor indices.

If the coupling between the system and environment is of the form
\begin{align}
\hat{H}_{SE}- \hat{H}_E = \hat{O}_S \otimes \hat{B}_E
\end{align}
then the environment correlation function is defined as 
\begin{align}
C(t_1,t_2) = \text{Tr}\left( \rho_E(0) \hat{B}_E(t_1) \hat{B}_E(t_2) \right)
\end{align}
where $\rho_E(0)$ is the state of the environment at time $t=0$.

For the typical case of a generic linear system-environment coupling $\hat{B}_E = \sum_k (g_k \hat{a}^\dagger_k + g^{*}_j \hat{a}_k)$, environment Hamiltonian $\hat{H}_E = \sum_k \omega_k \hat{a}^\dagger_k \hat{a}_k$, and initial environment state taken to be thermal with
inverse temperature $\beta$, the environment correlation function can be shown to be
\begin{align}
C(t) = \sum_k |g_k|^2 \left[ \coth(\beta \omega_k/2) \cos(\omega_k t) - i \sin(\omega_k t)\right] \ .
\end{align}

%

\end{document}